\documentclass[twocolumn]{aastex631}

\usepackage{amsmath}

\usepackage{hyperref}
\hypersetup{colorlinks=true, linkcolor=blue, citecolor=blue, filecolor=blue, urlcolor=blue}
\graphicspath{{./}{./plots/}}

\begin{document}

\title{A Unified Photometric Redshift Calibration for Weak Lensing Surveys using the Dark Energy Spectroscopic Instrument}
\shorttitle{DESI Redshift Calibration}

\correspondingauthor{Johannes U. Lange}
\email{E-mail: jlange@american.edu}
\shortauthors{J.~U.~Lange et al.}

\author[0000-0002-2450-1366]{J.~U.~Lange}
\affiliation{Department of Physics, American University, 4400 Massachusetts Avenue NW, Washington, DC 20016, USA}

\author{D.~Blanco}
\affiliation{Department of Astronomy and Astrophysics, UCO/Lick Observatory, University of California, 1156 High Street, Santa Cruz, CA 95064, USA}

\author[0000-0002-3677-3617]{A.~Leauthaud}
\affiliation{Department of Astronomy and Astrophysics, UCO/Lick Observatory, University of California, 1156 High Street, Santa Cruz, CA 95064, USA}
\affiliation{Department of Astronomy and Astrophysics, University of California, Santa Cruz, 1156 High Street, Santa Cruz, CA 95065, USA}

\author[0000-0001-7363-7932]{A.~H.~Wright}
\affiliation{Ruhr University Bochum, Faculty of Physics and Astronomy, Astronomical Institute (AIRUB), German Centre for Cosmological Lensing, 44780 Bochum, Germany}

\author[0009-0005-0530-0605]{A.~Fisher}
\affiliation{Department of Physics, American University, 4400 Massachusetts Avenue NW, Washington, DC 20016, USA}

\author{J.~Ratajczak}
\affiliation{Department of Physics and Astronomy, The University of Utah, 115 South 1400 East, Salt Lake City, UT 84112, USA}

\author{J.~Aguilar}
\affiliation{Lawrence Berkeley National Laboratory, 1 Cyclotron Road, Berkeley, CA 94720, USA}

\author[0000-0001-6098-7247]{S.~Ahlen}
\affiliation{Department of Physics, Boston University, 590 Commonwealth Avenue, Boston, MA 02215 USA}

\author[0000-0003-4162-6619]{S.~Bailey}
\affiliation{Lawrence Berkeley National Laboratory, 1 Cyclotron Road, Berkeley, CA 94720, USA}

\author[0000-0001-9712-0006]{D.~Bianchi}
\affiliation{Dipartimento di Fisica ``Aldo Pontremoli'', Universit\`a degli Studi di Milano, Via Celoria 16, I-20133 Milano, Italy}
\affiliation{INAF-Osservatorio Astronomico di Brera, Via Brera 28, 20122 Milano, Italy}

\author[0000-0002-5423-5919]{C.~Blake}
\affiliation{Centre for Astrophysics \& Supercomputing, Swinburne University of Technology, P.O. Box 218, Hawthorn, VIC 3122, Australia}

\author{D.~Brooks}
\affiliation{Department of Physics \& Astronomy, University College London, Gower Street, London, WC1E 6BT, UK}

\author{T.~Claybaugh}
\affiliation{Lawrence Berkeley National Laboratory, 1 Cyclotron Road, Berkeley, CA 94720, USA}

\author[0000-0002-2169-0595]{A.~Cuceu}
\affiliation{Lawrence Berkeley National Laboratory, 1 Cyclotron Road, Berkeley, CA 94720, USA}

\author[0000-0002-0553-3805]{K.~S.~Dawson}
\affiliation{Department of Physics and Astronomy, The University of Utah, 115 South 1400 East, Salt Lake City, UT 84112, USA}

\author[0000-0002-1769-1640]{A.~de la Macorra}
\affiliation{Instituto de F\'{\i}sica, Universidad Nacional Aut\'{o}noma de M\'{e}xico,  Circuito de la Investigaci\'{o}n Cient\'{\i}fica, Ciudad Universitaria, Cd. de M\'{e}xico  C.~P.~04510,  M\'{e}xico}

\author[0000-0002-0728-0960]{J.~DeRose}
\affiliation{Physics Department, Brookhaven National Laboratory, Upton, NY 11973, USA}

\author[0000-0002-4928-4003]{Arjun~Dey}
\affiliation{NSF NOIRLab, 950 N. Cherry Ave., Tucson, AZ 85719, USA}

\author{P.~Doel}
\affiliation{Department of Physics \& Astronomy, University College London, Gower Street, London, WC1E 6BT, UK}

\author{N.~Emas}
\affiliation{Centre for Astrophysics \& Supercomputing, Swinburne University of Technology, P.O. Box 218, Hawthorn, VIC 3122, Australia}

\author[0000-0003-4992-7854]{S.~Ferraro}
\affiliation{Lawrence Berkeley National Laboratory, 1 Cyclotron Road, Berkeley, CA 94720, USA}
\affiliation{University of California, Berkeley, 110 Sproul Hall \#5800 Berkeley, CA 94720, USA}

\author[0000-0002-3033-7312]{A.~Font-Ribera}
\affiliation{Institut de F\'{i}sica d’Altes Energies (IFAE), The Barcelona Institute of Science and Technology, Edifici Cn, Campus UAB, 08193, Bellaterra (Barcelona), Spain}

\author[0000-0002-2890-3725]{J.~E.~Forero-Romero}
\affiliation{Departamento de F\'isica, Universidad de los Andes, Cra. 1 No. 18A-10, Edificio Ip, CP 111711, Bogot\'a, Colombia}
\affiliation{Observatorio Astron\'omico, Universidad de los Andes, Cra. 1 No. 18A-10, Edificio H, CP 111711 Bogot\'a, Colombia}

\author[0000-0003-1481-4294]{C.~Garcia-Quintero}
\affiliation{Center for Astrophysics $|$ Harvard \& Smithsonian, 60 Garden Street, Cambridge, MA 02138, USA}

\author[0000-0001-9632-0815]{E.~Gaztañaga}
\affiliation{Institut d'Estudis Espacials de Catalunya (IEEC), c/ Esteve Terradas 1, Edifici RDIT, Campus PMT-UPC, 08860 Castelldefels, Spain}
\affiliation{Institute of Cosmology and Gravitation, University of Portsmouth, Dennis Sciama Building, Portsmouth, PO1 3FX, UK}
\affiliation{Institute of Space Sciences, ICE-CSIC, Campus UAB, Carrer de Can Magrans s/n, 08913 Bellaterra, Barcelona, Spain}

\author[0000-0003-3142-233X]{S.~Gontcho A Gontcho}
\affiliation{Lawrence Berkeley National Laboratory, 1 Cyclotron Road, Berkeley, CA 94720, USA}
\affiliation{University of Virginia, Department of Astronomy, Charlottesville, VA 22904, USA}

\author{G.~Gutierrez}
\affiliation{Fermi National Accelerator Laboratory, PO Box 500, Batavia, IL 60510, USA}

\author[0000-0002-7273-4076]{S.~Heydenreich}
\affiliation{Department of Astronomy and Astrophysics, UCO/Lick Observatory, University of California, 1156 High Street, Santa Cruz, CA 95064, USA}

\author[0000-0002-9814-3338]{H.~Hildebrandt}
\affiliation{Ruhr University Bochum, Faculty of Physics and Astronomy, Astronomical Institute (AIRUB), German Centre for Cosmological Lensing, 44780 Bochum, Germany}

\author[0000-0002-6024-466X]{M.~Ishak}
\affiliation{Department of Physics, The University of Texas at Dallas, 800 W. Campbell Rd., Richardson, TX 75080, USA}

\author[0000-0001-8528-3473]{J.~Jimenez}
\affiliation{Institut de F\'{i}sica d’Altes Energies (IFAE), The Barcelona Institute of Science and Technology, Edifici Cn, Campus UAB, 08193, Bellaterra (Barcelona), Spain}

\author[0000-0001-8820-673X]{S.~Joudaki}
\affiliation{CIEMAT, Avenida Complutense 40, E-28040 Madrid, Spain}

\author{R.~Kehoe}
\affiliation{Department of Physics, Southern Methodist University, 3215 Daniel Avenue, Dallas, TX 75275, USA}

\author[0000-0002-8828-5463]{D.~Kirkby}
\affiliation{Department of Physics and Astronomy, University of California, Irvine, 92697, USA}

\author[0000-0003-3510-7134]{T.~Kisner}
\affiliation{Lawrence Berkeley National Laboratory, 1 Cyclotron Road, Berkeley, CA 94720, USA}

\author[0000-0001-6356-7424]{A.~Kremin}
\affiliation{Lawrence Berkeley National Laboratory, 1 Cyclotron Road, Berkeley, CA 94720, USA}

\author{O.~Lahav}
\affiliation{Department of Physics \& Astronomy, University College London, Gower Street, London, WC1E 6BT, UK}

\author[0000-0002-6731-9329]{C.~Lamman}
\affiliation{Center for Astrophysics $|$ Harvard \& Smithsonian, 60 Garden Street, Cambridge, MA 02138, USA}

\author[0000-0003-1838-8528]{M.~Landriau}
\affiliation{Lawrence Berkeley National Laboratory, 1 Cyclotron Road, Berkeley, CA 94720, USA}

\author[0000-0001-7178-8868]{L.~Le~Guillou}
\affiliation{Sorbonne Universit\'{e}, CNRS/IN2P3, Laboratoire de Physique Nucl\'{e}aire et de Hautes Energies (LPNHE), FR-75005 Paris, France}

\author[0000-0003-1887-1018]{M.~E.~Levi}
\affiliation{Lawrence Berkeley National Laboratory, 1 Cyclotron Road, Berkeley, CA 94720, USA}

\author{L.~Medina-Varela}
\affiliation{Department of Physics, The University of Texas at Dallas, 800 W. Campbell Rd., Richardson, TX 75080, USA}

\author[0000-0002-1125-7384]{A.~Meisner}
\affiliation{NSF NOIRLab, 950 N. Cherry Ave., Tucson, AZ 85719, USA}

\author{R.~Miquel}
\affiliation{Instituci\'{o} Catalana de Recerca i Estudis Avan\c{c}ats, Passeig de Llu\'{\i}s Companys, 23, 08010 Barcelona, Spain}
\affiliation{Institut de F\'{i}sica d’Altes Energies (IFAE), The Barcelona Institute of Science and Technology, Edifici Cn, Campus UAB, 08193, Bellaterra (Barcelona), Spain}

\author[0000-0002-2733-4559]{J.~Moustakas}
\affiliation{Department of Physics and Astronomy, Siena College, 515 Loudon Road, Loudonville, NY 12211, USA}

\author[0000-0001-9070-3102]{S.~Nadathur}
\affiliation{Institute of Cosmology and Gravitation, University of Portsmouth, Dennis Sciama Building, Portsmouth, PO1 3FX, UK}

\author[0000-0001-8684-2222]{J.~ A.~Newman}
\affiliation{Department of Physics \& Astronomy and Pittsburgh Particle Physics, Astrophysics, and Cosmology Center (PITT PACC), University of Pittsburgh, 3941 O'Hara Street, Pittsburgh, PA 15260, USA}

\author[0000-0003-3188-784X]{N.~Palanque-Delabrouille}
\affiliation{IRFU, CEA, Universit\'{e} Paris-Saclay, F-91191 Gif-sur-Yvette, France}
\affiliation{Lawrence Berkeley National Laboratory, 1 Cyclotron Road, Berkeley, CA 94720, USA}

\author[0000-0002-2762-2024]{A.~Porredon}
\affiliation{CIEMAT, Avenida Complutense 40, E-28040 Madrid, Spain}
\affiliation{Institute for Astronomy, University of Edinburgh, Royal Observatory, Blackford Hill, Edinburgh EH9 3HJ, UK}
\affiliation{Ruhr University Bochum, Faculty of Physics and Astronomy, Astronomical Institute (AIRUB), German Centre for Cosmological Lensing, 44780 Bochum, Germany}
\affiliation{The Ohio State University, Columbus, 43210 OH, USA}

\author[0000-0001-7145-8674]{F.~Prada}
\affiliation{Instituto de Astrof\'{i}sica de Andaluc\'{i}a (CSIC), Glorieta de la Astronom\'{i}a, s/n, E-18008 Granada, Spain}

\author[0000-0001-6979-0125]{I.~P\'erez-R\`afols}
\affiliation{Departament de F\'isica, EEBE, Universitat Polit\`ecnica de Catalunya, c/Eduard Maristany 10, 08930 Barcelona, Spain}

\author{G.~Rossi}
\affiliation{Department of Physics and Astronomy, Sejong University, 209 Neungdong-ro, Gwangjin-gu, Seoul 05006, Republic of Korea}

\author[0000-0002-0394-0896]{R.~Ruggeri}
\affiliation{Queensland University of Technology,  School of Chemistry \& Physics, George St, Brisbane 4001, Australia}

\author[0000-0002-9646-8198]{E.~Sanchez}
\affiliation{CIEMAT, Avenida Complutense 40, E-28040 Madrid, Spain}

\author[0000-0002-0408-5633]{C.~Saulder}
\affiliation{Max Planck Institute for Extraterrestrial Physics, Gie\ss enbachstra\ss e 1, 85748 Garching, Germany}

\author{D.~Schlegel}
\affiliation{Lawrence Berkeley National Laboratory, 1 Cyclotron Road, Berkeley, CA 94720, USA}

\author{M.~Schubnell}
\affiliation{Department of Physics, University of Michigan, 450 Church Street, Ann Arbor, MI 48109, USA}
\affiliation{University of Michigan, 500 S. State Street, Ann Arbor, MI 48109, USA}

\author{D.~Sprayberry}
\affiliation{NSF NOIRLab, 950 N. Cherry Ave., Tucson, AZ 85719, USA}

\author[0000-0002-8246-7792]{Z.~Sun}
\affiliation{Department of Astronomy, Tsinghua University, 30 Shuangqing Road, Haidian District, Beijing, China, 100190}

\author[0000-0003-1704-0781]{G.~Tarl\'{e}}
\affiliation{University of Michigan, 500 S. State Street, Ann Arbor, MI 48109, USA}

\author{B.~A.~Weaver}
\affiliation{NSF NOIRLab, 950 N. Cherry Ave., Tucson, AZ 85719, USA}

\author[0000-0002-5992-7586]{S.~Yuan}
\affiliation{SLAC National Accelerator Laboratory, 2575 Sand Hill Road, Menlo Park, CA 94025, USA}

\author[0000-0002-7305-9578]{P.~Zarrouk}
\affiliation{Sorbonne Universit\'{e}, CNRS/IN2P3, Laboratoire de Physique Nucl\'{e}aire et de Hautes Energies (LPNHE), FR-75005 Paris, France}

\author[0000-0002-6684-3997]{H.~Zou}
\affiliation{National Astronomical Observatories, Chinese Academy of Sciences, A20 Datun Road, Chaoyang District, Beijing, 100101, P.~R.~China}

\begin{abstract}
    The effective redshift distribution $n(z)$ of galaxies is a critical component in the study of weak gravitational lensing. Here, we introduce a new method for determining $n(z)$ for weak lensing surveys based on high-quality redshifts and neural network-based importance weights. Additionally, we present the first unified photometric redshift calibration of the three leading stage-III weak lensing surveys, the Dark Energy Survey (DES), the Hyper Suprime-Cam (HSC) survey and the Kilo-Degree Survey (KiDS), with state-of-the-art spectroscopic data from the Dark Energy Spectroscopic Instrument (DESI). We verify our method using a new, data-driven approach and obtain $n(z)$ constraints with statistical uncertainties of order $\sigma_{\bar z} \sim 0.01$ and smaller. Our analysis is largely independent of previous photometric redshift calibrations and, thus, provides an important cross-check in light of recent cosmological tensions. Overall, we find excellent agreement with previously published results on the DES Y3 and HSC Y1 data sets while there are some differences on the mean redshift with respect to the previously published KiDS-1000 results. We attribute the latter to mismatches in photometric noise properties in the COSMOS field compared to the wider KiDS SOM-gold catalog. At the same time, the new $n(z)$ estimates for KiDS do not significantly change estimates of cosmic structure growth from cosmic shear. Finally, we discuss how our method can be applied to future weak lensing calibrations with DESI data.
\end{abstract}


\section{Introduction}

Since the first robust detection in the early 2000s \citep[see, e.g.,][]{Bacon2000_MNRAS_318_625, Kaiser2000_astro_ph_3338, VanWaerbeke2000_AA_358_30, Wittman2000_Natur_405_143}, weak gravitational lensing by large-scale structure has become a major cosmological probe. By studying the shape perturbations of the images of galaxies, we can obtain detailed measurements of the growth of structure over cosmic time. In the coming decade, experiments such as the Euclid satellite \citep{EuclidCollaboration2025_AA_697_1}, the Nancy Grace Roman Space Telescope \citep{Akeson2019_arXiv_1902_5569}, and the Vera C. Rubin Observatory \citep{TheLSSTDarkEnergyScienceCollaboration2018_arXiv_1809_1669} will image hundreds of millions of galaxies and detect weak gravitational lensing with unprecedented statistical precision. The increasing fidelity of these experiments puts stringent requirements on potential systematic errors, especially the calibration of photometric redshifts.

The observed weak lensing distortions of galaxy images by foreground structure depends not only on the foreground mass distribution but also on the distance to the lensed galaxies. As a result, translating weak lensing measurements into constraints on cosmic structure growth requires detailed knowledge about the distance to the lensed galaxies, i.e., their cosmological redshifts $z$. Precision redshifts for individual objects can, in principle, be obtained from spectroscopy. Unfortunately, given the large number of faint objects detected by weak lensing experiments, this approach is infeasible for all galaxies detected. Instead, weak lensing surveys rely on photometric redshifts whereby broad-band photometric galaxy colors, i.e., the ratio of fluxes in different bands of the electromagnetic spectrum, are used to estimate approximate, possibly biased redshifts and bin galaxies into different tomographic bins. Despite not knowing distances to individual galaxies in the different tomographic bins, unbiased cosmological constraints can be obtained as long as the redshift distribution $n(z)$ of galaxies in each tomographic bin is sufficiently well characterized.

To first order, weak lensing experiments are sensitive to the mean redshift $\bar z$ of each tomographic bin
\begin{equation}
    \bar z = \frac{\int_0^\infty z n(z) \mathrm{d}z}{\int_0^\infty n(z) \mathrm{d}z} \, .
\end{equation}
For stage-III experiments such as the Dark Energy Survey \citep[DES; ][]{Flaugher2015_AJ_150_150, Abbott2022_PhRvD_105_3520}, the Subaru Hyper Suprime-Cam Survey \citep[HSC; ][]{Aihara2018_PASJ_70_4, More2023_PhRvD_108_3520}, and the Kilo-Degree Survey \citep[KiDS; ][]{Kuijken2015_MNRAS_454_3500, Asgari2021_AA_645_104} the redshift distribution in each bin must be characterized to a precision of $\sigma_{\bar z} / (1 + \bar z) \sim 0.01$. For future stage-IV experiments such as the year-10 weak lensing analysis of the Vera C. Rubin Observatory, the requirement are as stringent as $\sigma_{\bar z} / (1 + \bar z) = 0.001$ \citep{TheLSSTDarkEnergyScienceCollaboration2018_arXiv_1809_1669}.

Different methods for the calibration of photometric redshifts have been developed \citep[see][and references therein]{Newman2022_ARAA_60_363}. One approach is to use high-quality, ideally spectroscopic, redshifts for a subset of weak lensing galaxies to determine $n(z)$ ``directly.'' A common difficulty with this so-called direct calibration is that the sample of galaxies with high-quality redshifts is typically not representative of the weak lensing sample. This requires additional steps to mitigate this bias, with self-organizing maps \citep[SOMs,][]{Kohonen2001_som_book} being a commonly used tool \citep{Masters2019_ApJ_877_81, Buchs2019_MNRAS_489_820, Wright2020_AA_637_100}. A SOM is an unsupervised machine-learning technique for dimensionality reduction. In the context of photometric redshifts, it is typically used to map the high-dimensional color space of galaxies into a two-dimensional representation whereby galaxies belonging to the same or nearby SOM cells have similar colors. This allows for the correction of non-representative samples of galaxies by comparing the distribution of SOM cells in the calibration sample to the wider weak lensing sample. A second widely-used approach is known as clustering redshifts. Most commonly, one uses angular cross-correlations between the weak lensing source galaxies and samples of objects with spectroscopic redshifts. The measured cross-correlation is proportional to $n(z) b(z)$, where $b(z)$ is the large-scale bias of the weak lensing sample. Assuming a reasonably flexible model for $b(z)$, one can then estimate $n(z)$.

The Dark Energy Spectroscopic Instrument \citep[DESI; ][]{Levi2013_arXiv_1308_0847, DESICollaboration2016_arXiv_1611_0036, DESICollaboration2016_arXiv_1611_0037, DESICollaboration2022_AJ_164_207, Silber2023_AJ_165_9, Miller2024_AJ_168_95, Poppett2024_AJ_168_245, DESICollaboration2024_AJ_167_62, DESICollaboration2024_AJ_168_58} is a stage-IV spectroscopic cosmology survey using a highly multiplexed fiber-fed spectrograph on the Mayall Telescope. DESI is able to measure spectroscopic redshifts for around $5000$ galaxies at a time \citep{Guy2023_AJ_165_144, Schlafly2023_AJ_166_259}. DESI, thus, is ideally poised to substantially improve photometric redshift calibration via both the direct and clustering redshift method. Recently, \cite{Ratajczak2025_arXiv_2508_9286} conducted a custom evaluation of all spectra obtained by DESI within two independent regions, one that encompasses the COSMOS field, and one that encompasses the XMM-LSS field. The targets for DESI spectroscopy in these two areas were selected for a broader range of analyses than the core large-scale structure samples, and thus represent a more representative sampling of the full galaxy population than the spectra over the full DESI footprint. At $r>16$, the surface density is $208\%$ higher than that for the DESI main sample for relatively red galaxies with $g-r>1$, and $250\%$ larger for relatively blue galaxies with color $g - r <1$. The sample of extended objects in an 7 $\mathrm{deg}^2$ annulus centered on the COSMOS field is roughly $24\%$ complete down to $i<23$.

In this work, we use the new DESI COSMOS-XMM catalog \citep{Ratajczak2025_arXiv_2508_9286} to perform an independent, unified photometric redshift calibration of the three leading stage-III weak lensing surveys, DES, HSC, and KiDS. A companion paper by Blanco et al. (in prep.) studies in more detail the potential for DESI to directly calibrate the KiDS data set. We show that DESI significantly increases the number of spectroscopic redshifts available in the COSMOS field. At the same time, DESI data does not fill the entire color space and misses redshifts above $z \sim 1.6$. In this paper, while we use the direct calibration method, we also introduce and verify a new technique that is based on importance weights instead of SOMs. Additionally, we compare our results against the fiducial calibration results obtained from the different weak lensing surveys. Our analysis method is substantially different from previous work and, additionally, we use a separate new catalog of spectroscopic redshifts from DESI, independent of the redshift compilations used in previous works. Thereby, our study addresses recent cosmological tensions from weak lensing surveys, most notably the $S_8$-tension \citep[see][and references therein]{Abdalla2022_JHEAp_34_49}. Finally, in a separate study, Ruggeri et al. (in prep.), we instead perform redshift calibrations with clustering redshifts using the DESI large-scale structure catalogs.

This paper is organized as follows. Section \ref{sec:data} briefly describes the different data sets used. In section \ref{sec:methods}, we describe our calibration method which is verified in section \ref{sec:verification}. Our main results, estimates of the redshift distributions $n(z)$ for DES, HSC, and KiDS, are presented in section \ref{sec:results}. We will discuss our results in section \ref{sec:discussion} before concluding in section \ref{sec:conclusion}. Throughout this work, we will assume cosmological parameters consistent with the \cite{PlanckCollaboration2020_AA_641_6} analysis of cosmic microwave background.

\section{Data}
\label{sec:data}

\begin{figure}
    \centering
    \includegraphics{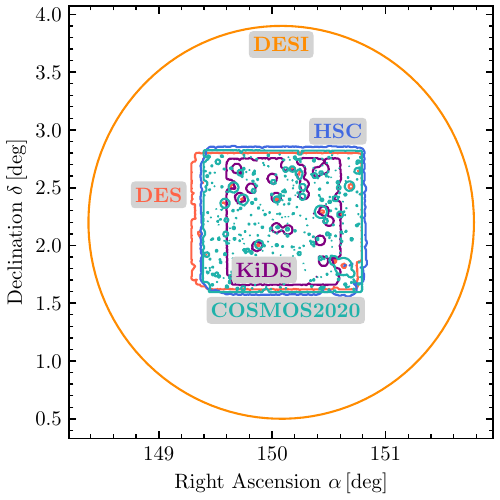}
    \caption{Approximate footprints of the different data sets used in this work in the COSMOS area.}
    \label{fig:map}
\end{figure}

Here, we describe our data, beginning with a redshift distribution calibration catalog of objects with high-quality redshifts followed by a description of data specific to the three lensing surveys. In Fig.~\ref{fig:map}, we show the footprints of the different data sets in the COSMOS field. Notably, the DESI catalog extends far beyond the COSMOS area for which we have deep lensing photometry, owing to DESI's large 1.6-degree radius focal plane.

\subsection{Calibration Data}

\begin{figure*}
    \centering
    \includegraphics{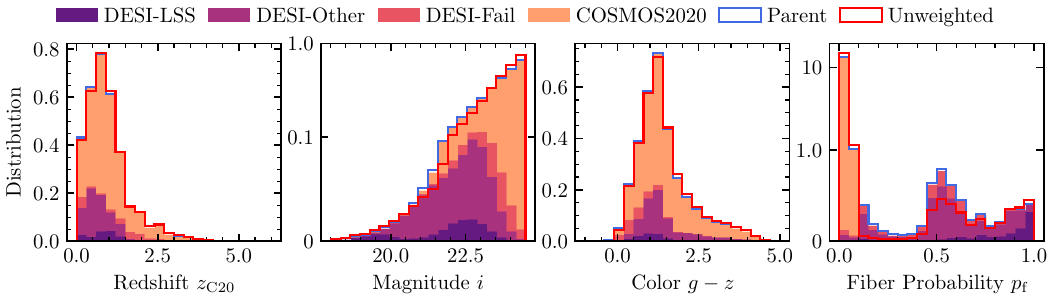}
    \caption{Properties of the calibration sample. From left to right, we show distributions in the COSMOS2020 redshifts $z_\mathrm{C20}$, the $i$ magnitude, the $g-z$ colors, and fiber probability $p_\mathrm{f}$ (right). In this figure, we limit our analysis to targets with $i < 24.5$. In all cases, we show the distribution of the COSMOS2020 parent sample as the blue line and the unweighted calibration sample as a red line. Finally, the colored histogram represents the weighted calibration sample and is divided in different target classes. These include targets with successful redshifts that are part of the BGS, LRG, ELG, and QSO samples (DESI-LSS), targets with DESI redshifts from other programs (DESI-Other), targets that received fibers from DESI but did not received a high-confidence redshift (DESI-Fail), and objects never targeted by DESI (COSMOS2020). Note that the $y$-axes of the middle and right histograms are partially logarithmic, i.e., beyond $0.1$ and $1.0$ for the middle and right histograms, respectively.}
    \label{fig:calibration}
\end{figure*}

Our calibration catalog is based on three distinct catalogs: the COSMOS2020 photometric catalog  \citep{Weaver2022_ApJS_258_11}, the deep photometric COSMOS catalog of the DESI Legacy Survey DR10 (Zhou et al., in prep.), and the spectroscopic DESI COSMOS-XMM catalog \citep{Ratajczak2025_arXiv_2508_9286}. The calibration catalog uses the COSMOS2020 catalog as a basis, adding in Legacy Survey and DESI information for individual objects where available. The deep Legacy Survey catalog in COSMOS has $5 \sigma$ imaging depths of $26.4$, $26.0$, $25.7$, and $24.9$ for the $griz$ bands. Thus, both COSMOS2020 and the Legacy Survey catalog are significantly deeper than the weak lensing catalogs considered here. The DESI COSMOS-XMM redshift catalogs include 307,934 unique spectra in total over the two $16$ deg$^2$ fields. This publicly available catalog includes redshift classifications and matches to the DESI Legacy Survey DR10 \citep{Dey2019_AJ_157_168}, HSC PDR3 \citep{Aihara2022_PASJ_74_247}, and COSMOS2020 \citep{Weaver2022_ApJS_258_11} photometry.

From COSMOS2020, we select all objects in the $2 \, \mathrm{deg}^2$ COSMOS area (\texttt{FLAG\_COMBINED==0}) with valid redshifts (\texttt{lp\_type!=-9}). We use the LePhare photometric redshifts, i.e., \texttt{lp\_zBEST}, for galaxies and stars and \texttt{lp\_zq} for X-ray sources. From the DESI COSMOS-XMM catalog, we keep all targets but consider a spectroscopic redshift unsuccessful if the redshift is not secure (\texttt{NOT QUALITY\_Z}) \citep{Ratajczak2025_arXiv_2508_9286}. We associate sources in the COSMOS2020 catalog with the DESI Legacy Survey DR10 catalog and the DESI catalog based on a $1 \, \mathrm{arcsec}$ matching radius. In case multiple exposures are present in the DESI catalog within a given matching radius, we give preference to successful redshifts in DESI's large-scale structure target classes, i.e., BGS, LRG, ELG, and QSO, and then other successful redshifts. Ultimately, each object in our calibration catalog will have COSMOS2020 photometry and photometric redshifts, a subset has Legacy Survey photometry, and an even smaller subset of that will have spectra, and possibly redshifts, from DESI.

In our analysis, we will use a subset of that calibration sample with the goal of extracting as much redshift information from the secure DESI spectroscopic redshifts. The basic idea is that galaxy populations with sufficient coverage by DESI will exclusively be calibrated with DESI redshifts. In Fig.~\ref{fig:calibration}, we display basic properties of the calibration sample such as redshift and magnitude distributions. We also demonstrate the weighting procedure that is described in more detail in section~\ref{subsec:incompleteness}.

\subsection{Dark Energy Survey}

The year 3 data release of the Dark Energy Survey (DES Y3) targets $\sim 4000 \, \mathrm{deg}^2$ on the sky with an effective number density of $5.6 \, \mathrm{arcmin}^{-2}$ \citep{Gatti2021_MNRAS_504_4312}. Galaxies are assigned into $4$ tomographic bins based on a SOM of the $3$-band $riz$ photometry \citep{Myles2021_MNRAS_505_4249}. In this work, we utilize the so-called Balrog catalogs \citep{Everett2022_ApJS_258_15} in which objects from deep photometric fields, such as COSMOS, are artificially injected into the DES Y3 footprint, including photometric noise. The artificial injections are processed with the same pipeline as the real data, allowing us to estimate the probability of each galaxy in the COSMOS field to pass the DES Y3 selection cuts and be assigned a certain tomographic bin as well as their lensing weight. Finally, the fiducial redshift calibration is based on a combination of SOM-based direct and clustering redshift calibration \citep{Myles2021_MNRAS_505_4249} and also accounts for the effects of blending \citep{MacCrann2022_MNRAS_509_3371}.

\subsection{Hyper Suprime-Cam}

The Hyper Suprime-Cam first-year data (HSC Y1), based on the intermediate data release S16a, covers $137 \, \mathrm{deg}^2$ with an effective number density of $\sim 22 \, \mathrm{arcmin}^{-2}$ \citep{Mandelbaum2018_PASJ_70_25}. Galaxies are assigned into $4$ tomographic bins based on photometric {\texttt Ephor AB} redshifts from $5$-band $grizy$ photometry \citep{Hikage2019_PASJ_71_43}. For our analysis, we make use of the S17A Wide-COSMOS catalog of the COSMOS field that is based on observations at the depth of the full survey, contains all objects passing the fiducial weak lensing selections cuts, and is matched against the COSMOS2015 catalog \citep{Laigle2016_ApJS_224_24}. In the default HSC first-year photometric redshift calibration, targets are weighted by the product of their lensing weights and a SOM-based weight to partially correct for cosmic variance in the COSMOS field. The fiducial $n(z)$ is then based on the weighted histogram of COSMOS2015 multi-band photometric redshifts, i.e., no spectroscopic redshifts were employed \citep{Hikage2019_PASJ_71_43}.

\subsection{Kilo-Degree Survey}

The KiDS-1000 data release of the Kilo-Degree Survey covers  $\sim 1000 \, \mathrm{deg}^2$ with an effective number density of $6.2 \, \mathrm{arcmin}^{-2}$ \citep{Giblin2021_AA_645_105}. A total of $5$ tomographic bins are assigned based on BPZ photometric redshifts from the $ugriZYJHK_s$ optical to near-infrared bands. For our analysis, we use a photometric catalog of the COSMOS field at a higher depth than the main survey that has been artificially degraded to simulate the depth of the KiDS-1000 weak lensing catalog (A. Wright, personal communication). Unlike for DES Y3 and HSC Y1, this KiDS-1000 COSMOS catalog has not been processed with the weak lensing pipeline. To estimate the impact of weak lensing cuts and weights, we also make use of the DR 4.1 photometric catalogs in combination with the KiDS-1000 weak lensing SOM-gold catalog, as described in section \ref{subsec:lensing_weights}. Finally, the fiducial photometric redshift calibration has been described in \cite{Hildebrandt2021_AA_647_124} and is based on a combination of SOM and clustering redshifts. 

\section{Methods}
\label{sec:methods}

In this section, we motivate and describe our method for photometric redshift calibration. The basic idea is to calculate two weights for each galaxy in the COSMOS field. The first one, a completeness weight, accounts for the fact that our calibration set is biased with respect to the general galaxy population. The second weight, a lensing weight, accounts for the probability that any galaxy in COSMOS makes the cut of a particular lensing survey and its expected weighting for the purpose of gravitational lensing studies. Finally, in this section, we also present estimates of statistical uncertainties associated with it.

\subsection{Incompleteness Correction}
\label{subsec:incompleteness}

While the DESI COSMOS spectroscopic catalog is extensive, not every galaxy detected photometrically in COSMOS has been spectroscopically observed with DESI, i.e., assigned a fiber. Furthermore, the sample of galaxies assigned fibers by DESI is based on cuts in photometry and, thus, is a biased subset of all galaxies in the COSMOS field. This bias can, to some extent, be corrected for with appropriate weighting. As a first step, for every galaxy in the Legacy Survey catalog, we estimate the fiber assignment probability, i.e., the probability that DESI attempted to obtain a spectroscopic redshift. Afterward, we determine a weight based on the fiber probability. This weight is designed to upweight galaxies that are unlikely to be assigned a fiber by DESI, i.e., those that are under-sampled.

In this work, we will use the following notation. The variable $F$ quantifies for every galaxy in the Legacy Survey whether it has been assigned a DESI fiber ($F = 1/\mathrm{True}$) or not ($F = 0/\mathrm{False}$). Similarly, we estimate a fiber fraction, i.e., the probability to be targeted for DESI spectra, for every galaxy and denote it by $p_\mathrm{f}$ ($0 \leq p_\mathrm{f} \leq 1$). For objects in the COSMOS2020 catalog that have no association in the Legacy Survey, we will always use $F = p_\mathrm{f} = 0$.

\subsubsection{Fiber Assignment Probabilities}
\label{subsec:fiber_probability}

Ideally, the fiber assignment probability would be computed from first principles if the targeting criteria were well-defined and we had random realizations of the fiber assignment such as for the DESI large-scale structure catalogs \citep{Lasker2025_JCAP_01_127}. In practice, the DESI COSMOS-XMM catalog contains targets from a large heterogeneous set of secondary target programs, often without alternative realizations of fiber assignments.

Instead, we approximate the probability $p_\mathrm{f}$ for each galaxy of being observed by DESI by comparing which galaxies in the photometric Legacy Survey COSMOS catalog were part of the spectroscopic DESI COSMOS-XMM catalog as a function of their photometry. Specifically, we employ fully connected multi-layer perceptron neural network as implemented in the \texttt{MLPClassifier} class in {\sc scikit-learn} version 1.5.2. Each network has $3$ hidden layers of $128$, $64$, and $32$ neurons, each. Furthermore, we employ the ReLU activation function, an L2 regularization of $10^{-4}$, early stopping, an optimization tolerance of $0$, a batch size of $100$, and an unlimited number of training iterations. All other hyperparameters were kept at their default values in {\sc scikit-learn}. The input values fed into the network are the sky coordinates, i.e., right ascension and declination, in addition to the $g$, $r$, $i$, and $z$ total and fiber fluxes, as well as the WISE W1 and W2 fluxes. We did not include flux errors as DESI targeting does not directly depend on them. The target values for the networks are whether any object in Legacy Survey was assigned a fiber by DESI, i.e., the variable $F$. The neural network hyperparameters described above were determined by optimizing the binary cross-entropy of the predicted fiber assignment probabilities in cross-validation analysis.

The values predicted by the trained network can be interpreted as the fiber fraction $p_\mathrm{f}$ given an object's coordinates and photometry. To avoid spurious overfitting we employ a K-fold approach by always training the networks on $90\%$ of the data and calculating $p_\mathrm{f}$ for the remaining $10\%$, i.e., $K=10$. Finally, this process is repeated $100$ times for the whole sample such that each individual galaxy has $100$ predictions. We typically find the scatter between different neural network predictions $\sigma_{p_\mathrm{f}}$ for the same galaxy to be of order $5\%$ or less. Thus, after averaging those predictions, the scatter is of order $0.5\%$ or less. We note that while the network has access to the coordinates of objects, it is unlikely to faithfully reproduce small-scale fiber assignment effects such as fiber collisions and the influence of higher-priority targets \citep{Lasker2025_JCAP_01_127}. This is not a concern for our work since we do not study small-scale clustering. Instead, we are mainly concerned with having accurate mean fiber probabilities as a function of photometry within overlap regions with weak lensing surveys. Finally, we observe that the raw $p_\mathrm{f}$ do not reproduce the mean fiber rate $f_\mathrm{f}$ in $5\%$ bins of $p_\mathrm{f}$ perfectly and instead show small offsets. We mitigate this by manually adjusting the raw $p_\mathrm{f}$ values via $p_\mathrm{f} \rightarrow f_\mathrm{f}(p_\mathrm{f})$.

\subsubsection{Importance Weights}

Naively, simply upweighting every galaxy in the DESI catalog by $p_\mathrm{f}^{-1}$ while disregarding galaxies not observed should result in a fair sample of galaxies in COSMOS. In practice, this does not work since some galaxies have a zero or near-zero fiber probability. If the fiber probability is zero, such targets cannot be recovered in a statistical sense even through upweighting. Furthermore, we do not expect estimates of $p_\mathrm{f}$ to be reliable for very small values since the relative uncertainty will be large. Finally, galaxies with $p_\mathrm{f} \sim 0$ would have very high weights $w$, leading to a small effective sample size $N_\mathrm{eff}$,
\begin{equation}
    N_\mathrm{eff} = \frac{\left( \sum w \right)^2}{\sum w^2} \, .
    \label{eq:N_eff}
\end{equation}
As described below, a small effective sample size would indicate a large statistical shot noise uncertainty in the effective redshift distribution. For these reasons, we instead use the following weighting scheme.
\begin{equation}
    w_\mathrm{c} =
    \begin{cases}
        p_\mathrm{f}^{-1} &\text{ if } p_\mathrm{f} > p_\mathrm{f, min} \text{ and } F \\
        0 &\text{ if } p_\mathrm{f} > p_\mathrm{f, min} \text{ and not } F \\
        1 &\text{ if } p_\mathrm{f} \leq p_\mathrm{f, min} \\
    \end{cases}
    \label{eq:weights}
\end{equation}
In the above equation, $0 \leq p_\mathrm{f, min} \leq 1$ is a free variable. In essence, targets with a fiber fraction above $p_\mathrm{f, min}$ only enter our sample if they received a fiber by DESI and, if they do, are upweighted by $p_\mathrm{f}^{-1}$. We note that this includes galaxies that were observed spectroscopically by DESI but did not obtain a successful, i.e., high-confidence redshift measurement. The rationale for this decision is that the DESI instrument is optimized for certain galaxies, e.g., luminous red galaxies or emission-line galaxies, at $z \lesssim 1.6$ \citep{DESICollaboration2016_arXiv_1611_0036}. Only using objects with high-confidence DESI redshifts would underestimate the number of galaxies at $z \gtrsim 1.6$ (Blanco et al., in prep.) and would likely lead to biases. This is issue is not limited to DESI as only using high-confidence spectroscopic redshifts from other surveys can lead to biases in weak lensing calibrations, as well \citep{Hartley2020_MNRAS_496_4769}. Finally, objects with a low fiber fraction enter our calibration sample regardless of whether they were observed by DESI or not, i.e., this includes objects that only have COSMOS2020 multi-band photometric redshifts. Those objects are not upweighted. Adjusting $p_\mathrm{f, min}$ changes how much of our sample is calibrated by DESI. Small values of $p_\mathrm{f, min}$ correspond to a large fraction of the sample having DESI spectra at the expense of larger shot noise and potential biases. Large values $p_\mathrm{f, min}$ will correspond to a lower fraction of the weight coming from galaxies observed with DESI while having a lower shot noise. In particular, $p_\mathrm{f, min} = 100 \%$ corresponds to taking the entire COSMOS2020 catalog without importance weighting, i.e., $w_\mathrm{c} \equiv 1$.

\subsection{Lensing Weights}
\label{subsec:lensing_weights}

In addition to using weights to correct for incompleteness in the DESI sample, we also need to apply weights related to the weak lensing analysis. In the following, we shall refer to a weak lensing sample $b$ as a certain combination of weak lensing catalog and tomographic bin, e.g., the second tomographic bin of DES. Three weights will need to be taken into account.

\begin{enumerate}
    \item Not every galaxy in the COSMOS field is expected to be part of a weak lensing sample $b$. To account for this, we weigh every galaxy by the probability $w_\mathrm{t}(b) = p(b)$, the probability for each galaxy to make the weak lensing quality cut and be assigned the tomographic bin associated with $b$. This is akin the so-called transfer function \citep{Myles2021_MNRAS_505_4249}.

    \item Galaxies in weak lensing surveys are typically assigned a weight $w_\mathrm{s}$ that is designed to minimize the variance of the measured shear. When calculating weak lensing statistics such as cosmic shear, galaxies are explicitly weighted by $w_\mathrm{s}$. Consequently, this weight needs to be taken into account when calculating the \textit{effective} redshift distribution $n(z)$ of weak lensing source galaxies.

    \item The measured ellipticity $e_\mathrm{obs}$ of galaxies may have a non-unity response to changes in the intrinsic ellipticity $e_\mathrm{int}$ induced by gravitational shear, i.e., $e_\mathrm{obs} = R \times e_\mathrm{int}$ with $R \neq 1$. This also affects the effective redshift distribution, essentially weighting each galaxy by $w_R = R$.
\end{enumerate}

The total lensing weight of each galaxy with respect to the lensing sample $b$ is thus 
\begin{equation}
    w_\mathrm{l}(b) = w_\mathrm{t}(b) \langle w_\mathrm{s} w_R \rangle_b \, ,
\end{equation}
where $\langle \rangle_b$ denotes the expectation value over random realizations of the weak lensing survey where the galaxy ends up in sample $b$. In the following, we will describe how these weights are taken into account in practice for the different weak lensing surveys.

For every galaxy in the COSMOS field, the DES Balrog catalog \citep{Everett2022_ApJS_258_15} contains a large number of artificial injections $n_\mathrm{inj}$ throughout the DES survey footprint. We use these injections to estimate $p(b)$, i.e., $n_b / n_\mathrm{inj}$, where $n_b$ is the number of injections that fall inside $b$. Similarly, we use the Balrog catalog to estimate $\langle w_\mathrm{s} w_R \rangle_{b}$ through a simple Monte-Carlo estimate. For DES, the response $R$ is estimated through the diagonal of the response matrix, $0.5 (R_{11} + R_{22})$ \citep{Huff2017_arXiv_1702_2600}. Unlike for DES, artificial injections of COSMOS galaxies into the weak lensing survey footprint do not exist for HSC. Instead, the COSMOS area has been observed with conditions expected for the wider HSC weak lensing survey. For every COSMOS galaxy, we thus only have a single realization within the wider weak lensing catalog. Thus, $p(b)$ is $1$ if the galaxy passed the lensing cut and is assigned a certain redshift bin and $0$, otherwise. Similarly, $\langle w_\mathrm{s} w_R \rangle_b$ becomes $w_\mathrm{s} w_R$ and since the shear response is not available for HSC, we just use $w_\mathrm{s}$. Finally, the weights for KiDS are estimated in a similar fashion as for HSC. However, unlike for HSC, no lensing cuts or weights have been determined for galaxies in the COSMOS field. Thus, we learn the relation between photometry in the the raw KiDS DR4 photometric catalog and the total lensing weight $w_\mathrm{l}$ in the KiDS-1000 SOM-gold catalog using neural networks, i.e., with the \texttt{MLPRegressor} class in {\sc scikit-learn}, a $\tanh$ activation function, early stopping, no L2 regularization, a batch size of $10,000$, an optimization tolerance of $0$ and $3$ hidden layers of $128$, $64$, and $32$ neurons, each.

\subsection{Redshift Calibration}

To obtain an estimate of the redshift distribution $n(z)$ of a given weak lensing sample $b$, we start by spatially matching the weak lensing sample in COSMOS against the COSMOS2020-DESI calibration catalog using a $1$ arcsec matching radius. We only use galaxies that are present in both the weak lensing catalog and the calibration catalog. Since the COSMOS2020 catalog is much deeper than the weak lensing surveys, only using a subset of the weak lensing catalog present in the calibration catalog should not result in systematic biases with respect to photometry. The final weight given to each galaxy is the product of the completeness weight in the calibration catalog and the weak lensing weight,
\begin{equation}
    w_\mathrm{tot} = w_\mathrm{c} w_\mathrm{l}(b) \, .
\end{equation}
The high-quality redshift assigned to each galaxy is the DESI redshift, if available, and the COSMOS2020 multi-band photometric redshift, otherwise. In the end, $n(z)$ is estimated from a simple weighted histogram of the high-quality redshifts.

\subsection{Statistical Uncertainties}

Our redshift calibration is affected by two sources of statistical uncertainty. First, we probe the cosmological volume in our calibration field with a finite number of galaxies, leading to shot noise (SN). Furthermore, matter and galaxies are spatially clustered. As a result, in our calibration fields, there may be matter and galaxy overdensities at certain redshifts. This uncertainty, which cannot be reduced by increasing the sampling density, will be referred to as cosmic variance (CV). We approximate both sources of uncertainty as uncorrelated such that, for example, if we measure the mean effective redshift $\bar z$ of a weak lensing catalog, we expect
\begin{equation}
    \sigma_{\bar z}^2 = \sigma_{\bar z, \mathrm{SN}}^2 + \sigma_{\bar z, \mathrm{CV}}^2 \, .
\end{equation}

We note that our photometric redshift calibration may also be affected by other systematic uncertainties such as the photometric calibration in the calibration fields as well as errors in the redshifts of galaxies in our calibration sample. We will return to these systematic uncertainties in section \ref{sec:discussion}.

\subsubsection{Shot Noise}

We probe the cosmological volume in our calibration field with a finite but sufficiently large sample of galaxies. Furthermore, galaxies in our calibration sample have non-uniform weights $w_\mathrm{tot}= w$. Given the large sample size, a good approximation to the standard error on the mean redshift is then
\begin{equation}
    \sigma_{\bar z, \mathrm{SN}}^2 \approx \frac{\sum w_i^2 (\bar z - z_i)^2}{\sum w_i^2} = \frac{\sigma_z^2}{N_\mathrm{eff}},
    \label{eq:shot_noise}
\end{equation}
where $\sigma_z^2$ is an estimate of the variance of redshift distribution in the weak lensing catalog and $N_\mathrm{eff}$ the effective sample size,
\begin{equation}
    N_\mathrm{eff} = \frac{\left( \sum w_i \right)^2}{\sum w_i^2} \frac{\sum w_i (z_i - \bar z)^2}{\sum w_i} \frac{\sum w_i^2}{\sum w_i^2 (z_i - \bar z)^2} \, .
\end{equation}
Additionally, when comparing two samples of the same galaxies but different weights, $w_1$ and $w_2$, an estimate of the correlation in the mean redshift is
\begin{equation}
    \rho_{\mathrm{SN}, \bar{z}_1, \bar{z}_2} = \frac{\sum\limits w_{1, i} w_{2, i} (z - \bar z)^2}{\sqrt{\sum w_{1, i}^2 (z - \bar z)^2 \sum w_{2, i}^2 (z - \bar z)^2}} \, .
    \label{eq:d_z_sn}
\end{equation}

\subsubsection{Cosmic Variance}

Matter and galaxies are spatially clustered, forming over- and underdensities on large scales. As a result, certain redshifts may be over- or underrepresented in our calibration sample in a way that is independent of shot noise \citep{Moster2011_ApJ_731_113}. To account for this, we first calculate the covariance $C_{ij}$ in the relative matter overdensity between two volumes at redshifts $z_i$ and $z_j$. The matter variance depends on the matter correlation function $\xi(r, z)$, where $r$ is the comoving distance between two points \citep{Moster2011_ApJ_731_113},
\begin{equation}
    \begin{split}
        C_{ij} =& \frac{1}{V_i V_j} \int \int \xi \left( r_{ij}, z \right) \mathrm{d}V_i \mathrm{d}V_j \\
        \approx& \frac{1}{\Delta d_i \Delta d_j} \int \int \int f(\theta) \\
        &\xi \left( \sqrt{\tilde{d}_{ij}^2 \theta^2 + (d_i - d_j)^2}, z \right) \mathrm{d}d_i \mathrm{d}d_j \mathrm{d} \theta \, .
    \end{split}
\end{equation}
In the above equation, $d$ refers to the comoving distance, $\tilde{d}_{ij} = 0.5 (d_i + d_j)$ and $f(\theta)$ is the normalized distribution of separation angles between two random points inside the survey footprint. The distribution $f(\theta)$ can be directly estimated from the COSMOS calibration catalog assuming that galaxies are only weakly clustered. Using {\sc CAMB} \citep{Lewis2000_ApJ_538_473, Howlett2012_JCAP_04_027} version 1.5.5 to calculate the linear matter power spectrum at different cosmic times, we can then calculate the matter covariance between different redshifts $z_i$ and $z_j$.

Galaxies are biased tracers of the underlying matter field such that the galaxy covariance between $z_i$ and $z_j$ is $C_{ij} b_i b_j$, where $b_i$ and $b_j$ are the linear galaxy bias parameters at $z_i$ and $z_j$, respectively. Ultimately, one can show that the cosmic variance term can be approximated as
\begin{equation}
    \sigma_{\bar z, \mathrm{CV}}^2 = \frac{\int n(z_i) n(z_j) (z_i - \bar z) (z_j - \bar z) b_i b_j C_{ij} \mathrm{d} z_i \mathrm{d} z_j}{\left( \int n(z) \mathrm{d} z \right)^2} \, .
\end{equation}
Note that the galaxy bias is unknown but we can reasonably expect $b \sim 1 - 2$. In the following, we will assume a constant, redshift-independent galaxy bias $b$ and express our results in $\sigma_{\bar z, \mathrm{CV}}^2 / b^2$.

\section{Verification}
\label{sec:verification}

Before determining effective redshift distributions for the different weak lensing samples, we want to verify that our method works reliably.

\subsection{Fiber Assignment Probabilities}

\begin{figure*}[h!]
    \centering
    \includegraphics{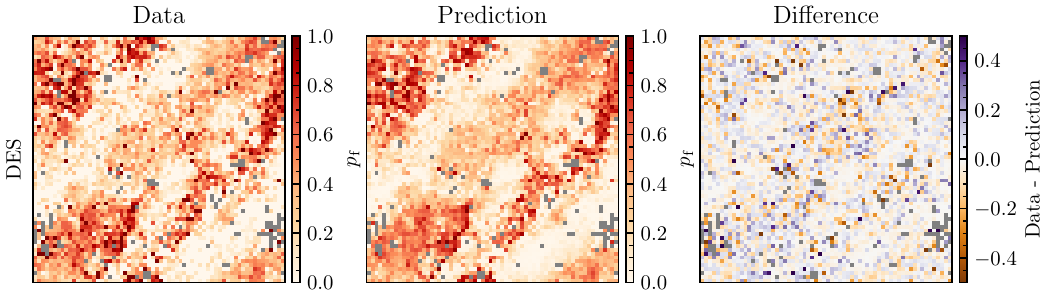}
    \includegraphics{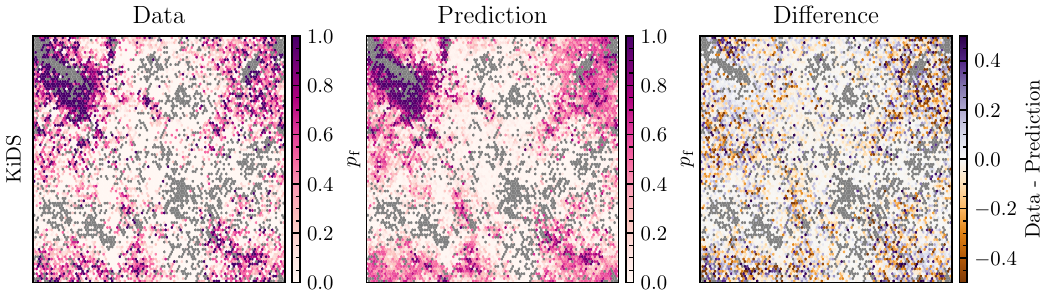}
    \caption{Visual check of the accuracy of our machine learning model of the fiber fraction $p_\mathrm{f}$. We show galaxies in the DES (top row) and KiDS (bottom row) COSMOS sample projected onto the DES and KiDS SOM, respectively. The left panel shows the fraction of galaxies in each SOM cell observed by DESI. The middle panel shows the predicted fraction in each cell, i.e., the mean predicted fiber fraction of all galaxies in each cell. Finally, the right panel shows the difference between the left and middle panel. Overall, our machine learning model for $p_\mathrm{f}$ reproduces the trends as a function of DES and KiDS photometry seen in the data. Grey cells indicate SOM cells not present in the final weak lensing catalogs.}
    \label{fig:p_fiber}
\end{figure*}

A key ingredient of our analysis is an accurate estimate of the DESI fiber probability of galaxies in the COSMOS field. As a first step, we perform a visual inspection via SOMs to verify that the results predicted by the deep learning model are plausible. We use SOMs as compared to, e.g., single-band magnitudes or two-band colors, because they allow us to study fiber probabilities in high-dimensional color and magnitude space.\footnote{We also studied the $grizy$ magnitude, color, and color-color distributions of DES/KiDS targets matched to COSMOS2020, similar, to Fig. \ref{fig:calibration}. We compared distributions with and without importance weights and found them to be in very good agreement.} In Fig.~\ref{fig:p_fiber}, we show the completeness in DESI fiber assignment for galaxies in the DES and KiDS COSMOS catalogs. For DES, galaxies are projected onto the deep DES SOM \citep{Myles2021_MNRAS_505_4249} and KiDS galaxies are projected onto the KiDS SOM \citep{Wright2020_AA_637_100}. The left-hand panel shows the actual fiber fraction in the SOM whereas the middle panel shows the predicted completeness. The predicted completeness in each SOM cell is simply the average $p_\mathrm{f}$ of all galaxies in each cell. It is worth emphasizing that these are pure predictions. As described in section \ref{subsec:fiber_probability}, we used a K-fold approach. As a result, the $p_\mathrm{f}$ estimates for each galaxy are never obtained from a model that was trained on the galaxy itself. Overall, the network correctly predicts the trends seen in the data. The right-hand panel, showing the difference between data and model predictions, shows no strong trends with color. To evaluate this agreement more quantitatively, we calculate the coefficient of determination $R^2$ for the predicted and observed fiber fractions within each SOM cell. We find $R^2$ to be $0.816$ and $0.665$ for DES and KiDS, respectively. From Monte-Carlo simulations under the assumption that the fiber fractions are accurate and all fiber assignments are independent of each other, we find that the expected values are $0.813 \pm 0.008$ and $0.691 \pm 0.007$, respectively. The agreement is very good for DES but the correlation is slightly weaker than expected for KiDS. This may indicate slight inaccuracies in the estimated fiber probabilities but, based on the tests performed in section \ref{subsec:cosmos_redshifts} below, we do not think this has a substantial impact on the inferred redshift distributions.

\subsection{Impact of Fiber Fraction Threshold}

\begin{figure*}
    \centering
    \includegraphics{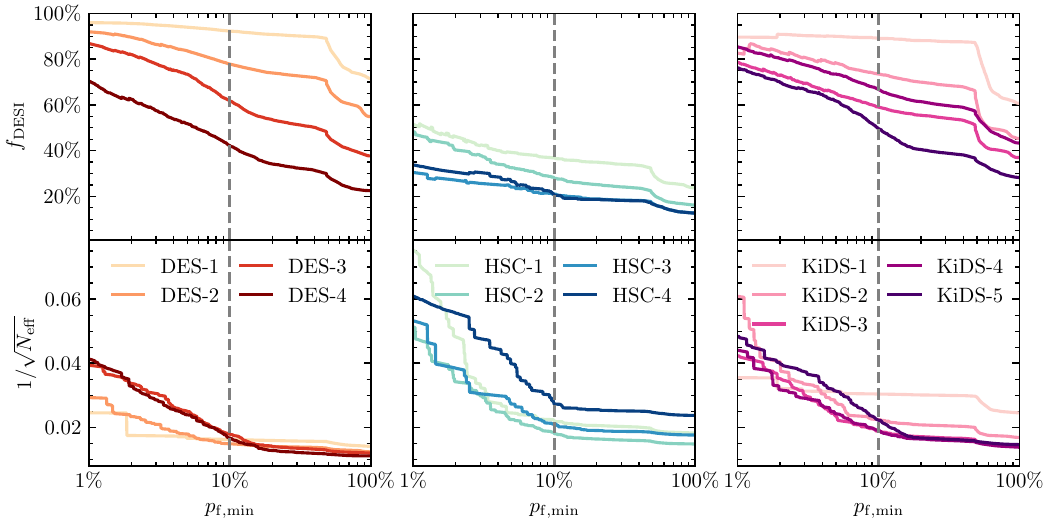}
    \caption{Impact of the choice of $p_\mathrm{f, min}$ on the fraction of the weak lensing sample calibrated with DESI redshifts and shot noise. Each column corresponds to one of the three weak lensing samples: DES, HSC, and KiDS. Upper panels show the weighted fraction of galaxies in the calibration sample with successful redshifts from DESI. Note that galaxies observed with DESI but without a successful redshift are counted as not having a DESI redshift, i.e., they will drive $f_\mathrm{DESI}$ down. The lower panel shows the inverse square root of the effective sample size, a measure of the shot noise uncertainty in the sample. The grey vertical line in the upper panels indicate our default choice, $p_\mathrm{f, min} = 0.10$.}
    \label{fig:f_desi}
\end{figure*}

The fiber fraction threshold $p_\mathrm{f, min}$ is a key parameter in our method that effectively determines how strongly the redshift calibration relies on galaxies observed by DESI as opposed to galaxies with high-quality multi-band photometric redshifts from COSMOS2020\footnote{A similar concept, \texttt{Pphot}, has been used by \cite{Speagle2019_MNRAS_490_5658} to study and adjust the reliance of the HSC calibration on many-band photometric redshifts as opposed to spectroscopic redshifts.}. In Fig.~\ref{fig:f_desi}, we study how changes in $p_\mathrm{f, min}$ affect the relative contribution of DESI redshifts and shot noise for DES, HSC, and KiDS.

Top panels indicate the weighted average of galaxies with DESI redshifts where the weight is $w_\mathrm{tot}$. Note that galaxies observed by DESI that did not end up getting a spectroscopic redshift will not be counted in this fraction. As expected, for all weak lensing samples, reducing $p_\mathrm{f, min}$ leads to an increasing fraction of galaxies with DESI redshifts. Furthermore, we see that the shallower DES and KiDS surveys can be calibrated with a much higher fraction of DESI galaxies than the deeper HSC survey. Similarly, higher-redshift tomographic bins receive less contributions from galaxies with DESI redshifts. Bottom panels indicate the amount of shot noise as measured by the effective sample size defined in equation \eqref{eq:N_eff}. Unsurprisingly, lowering $p_\mathrm{f, min}$, corresponding to using fewer galaxies with only high-quality photometric redshifts, leads to an increase in the shot noise. The increase is especially strong for $p_\mathrm{f, min} < 0.1$.

\subsection{Redshift Distribution}
\label{subsec:cosmos_redshifts}

The main goal of this analysis is an accurate recovery of the effective redshift distribution $n(z)$ of different photometric weak lensing samples. Most photometric redshift calibration studies employ mock data from simulations, such as MICE \citep{Carretero2015_MNRAS_447_646} and Buzzard \citep{DeRose2019_arXiv_1901_2401}, to help with this task, particularly when it comes to verifying the spectroscopic incompleteness correction. However, in this work, we argue that rigorous tests can also be performed on the data itself. The main idea is that high-quality COSMOS2020 photometric redshifts $z_\mathrm{C20}$ exist for most galaxies in the COSMOS field. Thus, for every weak lensing catalog, we can directly measure the $z_\mathrm{C20}$ redshift distribution in the COSMOS field and test how well we can recover this distribution from a biased subset of galaxies with high-quality redshifts. While COSMOS2020 redshifts have a certain scatter and catastrophic outliers around the true (spectroscopic) redshift \citep{Weaver2022_ApJS_258_11}, especially for faint targets observed with HSC, the correlation with the true redshift is very strong such that one should expect that biases in the recovered $n(z)$ should show up in $n(z_\mathrm{C20})$.

\begin{figure*}
    \centering
    \includegraphics{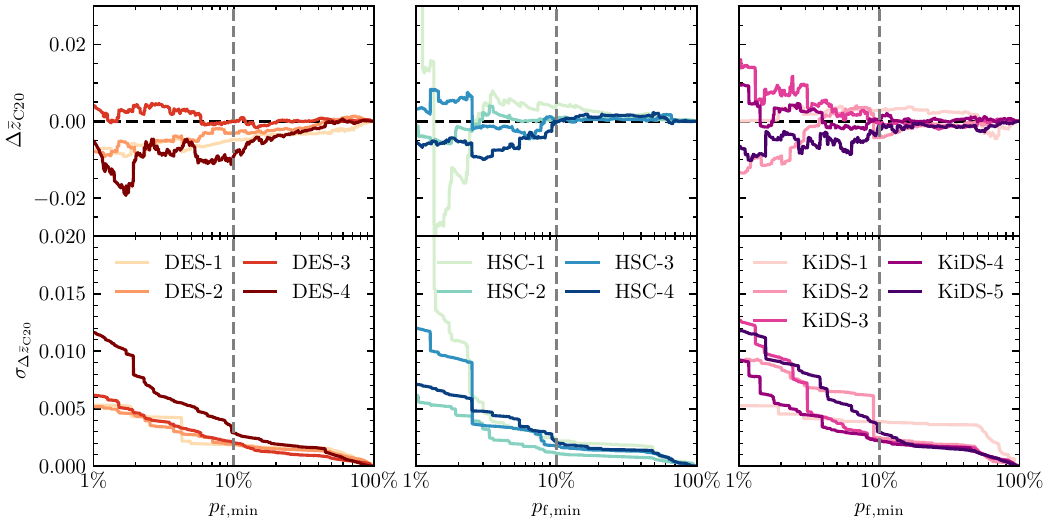}
    \caption{Verification of the importance weighting method with COSMOS2020 photometric redshifts. We show the difference in the weighted-average COSMOS2020 redshift as a function of $p_\mathrm{f, min}$ with respect to $p_\mathrm{f, min} = 100 \%$. As disussed in the text, $p_\mathrm{f, min} = 100 \%$ corresponds to an unweighted sample of all galaxies in the COSMOS2020 catalog. Each column corresponds to one of the three weak lensing samples: DES, HSC, and KiDS. Upper panels show the absolute difference whereas lower panels shows an estimate of the significance of the difference after accounting for shot noise.}
    \label{fig:d_z}
\end{figure*}

In order to check for biases in $n(z_\mathrm{C20})$, we modify our pipeline to always use $z_\mathrm{C20}$ for all galaxies, even if a DESI redshift is available. As discussed earlier, setting $p_\mathrm{f, min} = 1$ corresponds to no upweighting and, thus, gives an unbiased estimate of $n(z_\mathrm{C20})$. Thus, we check whether $n(z_\mathrm{C20})$ changes in a statistically significant manner as we lower $p_\mathrm{f, min}$. Since we are primarily interested in the mean effective redshift, the top panels of Fig.~\ref{fig:d_z} shows the change in the mean $z_\mathrm{C20}$ redshift $\bar{z}_\mathrm{C20} = \int z_\mathrm{C20} n(z_\mathrm{C20}) \mathrm{d}z_\mathrm{C20}$ compared to $p_\mathrm{f, min} = 1$. The lower panels of the same figure show an estimate of the statistical uncertainty in that difference, as calculated via eq.~\eqref{eq:d_z_sn}.

Overall, we see that the shifts in the mean $z_\mathrm{C20}$ redshifts are very small, staying well below $\Delta \bar{z}_\mathrm{C20} < 0.005$ as long as $p_\mathrm{f, min} > 0.10$. Thus, in the following we choose $p_\mathrm{f, min} = 0.10$ in our default analysis since it provides a good compromise between accuracy, shot noise, and a high fraction of the calibration coming from DESI redshifts. We note that for current stage-III lensing surveys, mean redshifts should be calibrated to within $\sigma_{\bar z} / (1 + \bar z) \sim 0.01$. The shifts we observe stay well below that limit and approach the precision needed for stage-IV surveys.

Finally, we note that the test performed here is primarily a test related to the recovery of intrinsic redshift distributions from biased samples, similar to most tests with mock catalogs \citep[see, e.g.,][]{Wright2020_AA_637_100, Myles2021_MNRAS_505_4249}. It is worth reiterating that for our fiducial analysis, we use COSMOS2020 redshifts which have scatter around true redshifts and may bias our estimates of $n(z)$. The same is true, but to a lesser extent, for redshifts from DESI. Similarly, we do not estimate potential biases in the photometric calibration in the deep fields \citep[see, e.g.,][]{Myles2021_MNRAS_505_4249}. We leave such investigations to future work.

\section{Results}
\label{sec:results}

\begin{figure*}
    \centering
    \includegraphics{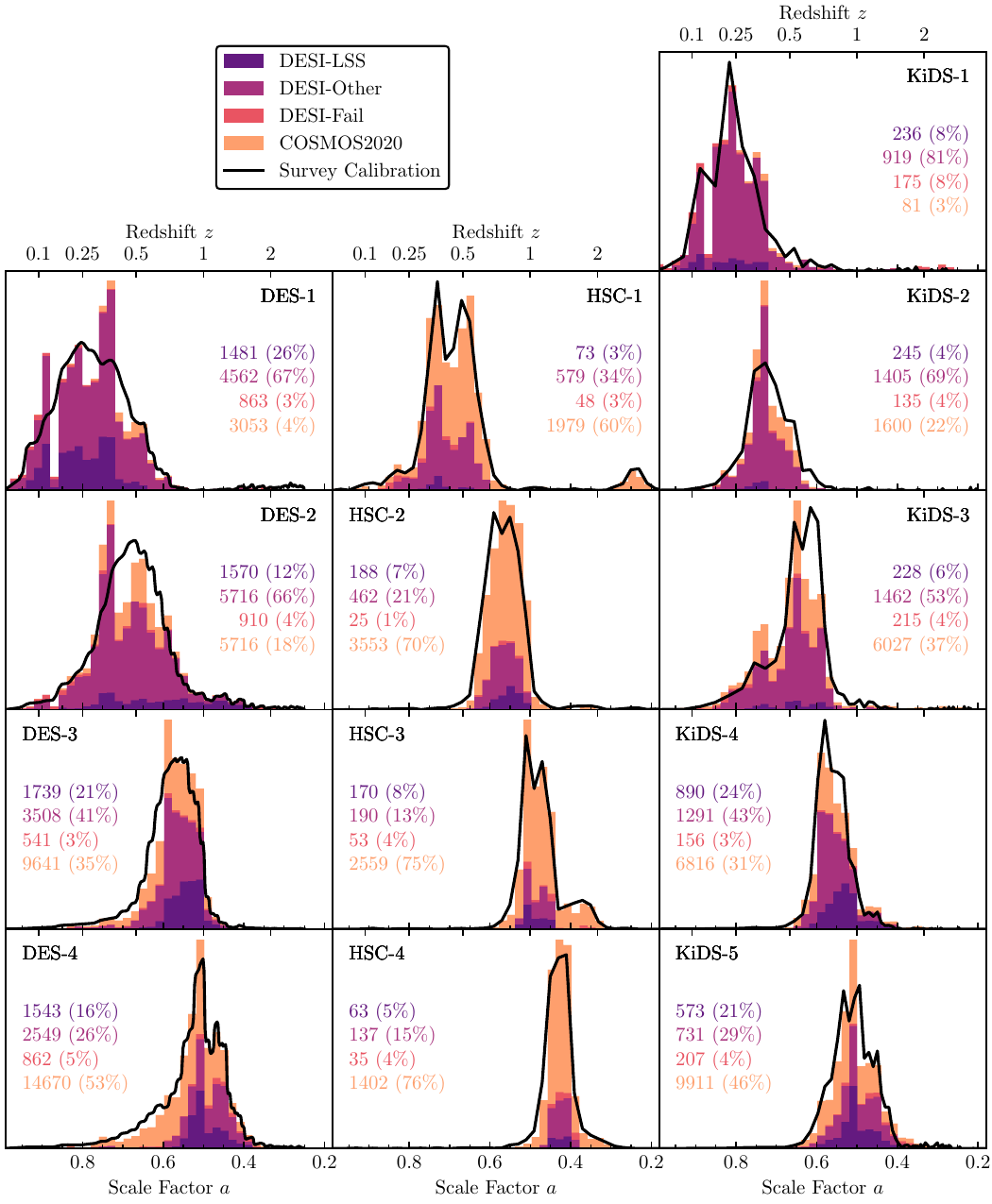}
    \caption{Estimates of the effective redshift distributions of source galaxies in different tomographic bins for DES, HSC, and KiDS. The colored filled histogram shows our estimates, broken down into different redshift types, similar to Fig.~\ref{fig:calibration}. The numbers in each panel denote the absolute number of galaxies in each category and their fractional weighted contribution to the $n(z)$ estimate. Finally, the calibration determined by the respective weak lensing collaboration is also shown as a black line for comparison.}
    \label{fig:n_z}
\end{figure*}

\begin{table}
\centering
\caption{Photometric redshift calibration results for the default parameter value, $p_\mathrm{f, min} = 0.10$.}
\begin{tabular}{cccccc}
\hline
Sample & $\bar{z}$ & $0.01 \times (1 + \bar{z})$ & $\sigma_{\bar z, \mathrm{SN}}$ & $\sigma_{\bar z, \mathrm{CV}} / b$ & $\Delta \bar{z}_\mathrm{fid}$ \\
\hline\hline
DES-1 & $0.322$ & $0.013$ & $0.003$ & $0.012$ & $-0.014$ \\
DES-2 & $0.513$ & $0.015$ & $0.003$ & $0.009$ & $-0.009$ \\
DES-3 & $0.758$ & $0.018$ & $0.003$ & $0.005$ & $+0.017$ \\
DES-4 & $0.940$ & $0.019$ & $0.004$ & $0.005$ & $+0.004$ \\
HSC-1 & $0.550$ & $0.016$ & $0.010$ & $0.010$ & $-0.015$ \\
HSC-2 & $0.807$ & $0.018$ & $0.004$ & $0.004$ & $+0.014$ \\
HSC-3 & $1.131$ & $0.021$ & $0.005$ & $0.005$ & $+0.004$ \\
HSC-4 & $1.357$ & $0.024$ & $0.005$ & $0.004$ & $+0.011$ \\
KiDS-1 & $0.293$ & $0.013$ & $0.008$ & $0.012$ & $+0.036$ \\
KiDS-2 & $0.395$ & $0.014$ & $0.006$ & $0.005$ & $-0.006$ \\
KiDS-3 & $0.544$ & $0.015$ & $0.004$ & $0.007$ & $-0.007$ \\
KiDS-4 & $0.819$ & $0.018$ & $0.003$ & $0.005$ & $+0.038$ \\
KiDS-5 & $1.006$ & $0.020$ & $0.005$ & $0.005$ & $+0.016$ \\
\hline
\end{tabular}
\tablecomments{The rightmost column shows the difference with respect to the fiducial results published by the weak lensing collaborations.}
\label{tab:n_z}
\end{table}

In Fig.~\ref{fig:n_z}, we show the main result of our study: estimates of the effective redshift distributions for DES, HSC, and KiDS. We also compare these results against the published $n(z)$ of the three weak lensing survey \citep{Hikage2019_PASJ_71_43, Hildebrandt2021_AA_647_124, Myles2021_MNRAS_505_4249}. The results are plotted as $n(a)$, where $a = (1+z)^{-1}$ is the scale factor, instead of $n(z)$ for visual clarity. For convenience, we also list the results for the mean redshift in Table~\ref{tab:n_z}, including estimates of shot noise and cosmic variance, as well as the difference compared to the published results. For KiDS-1000, the fiducial results take into account the bias as measured from the MICE mocks \citep[Col.~2, Table~3 in][]{Hildebrandt2021_AA_647_124}.

\subsection{Dark Energy Survey}

Overall, we find good agreement with the results from DES \citep{Myles2021_MNRAS_505_4249}, both in terms of the mean and distribution of redshifts. The strongest difference in the mean redshift occurs for the third tomographic bin, DES-3, with $\Delta \bar z_\mathrm{fid} = +0.017$. However, the published DES Y3 results incorporate the effects of blending \citep{MacCrann2022_MNRAS_509_3371} as well as additional constraints from clustering redshifts \citep{Gatti2022_MNRAS_510_1223}. When comparing to the DES Y3 results without these effects, the difference between our and the fiducial results reduces further with $\Delta \bar z_\mathrm{fid} = [-0.010, -0.007, +0.008, -0.004]$ for the 4 tomographic bins \citep[see Table 2 in][]{Myles2021_MNRAS_505_4249}.

Furthermore, for DES Y3, we can estimate whether the redshift distribution in the overlap region with the DESI main survey footprint, i.e., at declination $\delta > -10 \, \mathrm{deg}$, is significantly different from the distribution in the full DES Y3 footprint. To do so, we recompute the lensing weights, $w_R$, using only Balrog injections at $\delta > -10 \, \mathrm{deg}$. We find shift towards slightly lower mean redshifts, i.e., $\Delta \bar = -0.002 \pm 0.003$, $-0.008 \pm 0.003$, $-0.009 \pm 0.003$, and $-0.009 \pm 0.005$ for the four tomographic bins. This suggest that using the full-footprint redshift distributions in DESI galaxy-galaxy lensing studies does not lead to significant biases in the inferred cosmology.

\subsection{Hyper Suprime-Cam}

The results of this work agree well with the published HSC results. That is expected, to some extent, since the HSC study is based on a direct calibration with COSMOS2015 multi-band photometric redshifts. While we also use DESI spectroscopic redshifts, the HSC results, due to the deeper depth, still rely largely on updated COSMOS2020 multi-band photometric redshifts, as shown in Fig.~\ref{fig:n_z}. Nonetheless, this result also demonstrates a good agreement of SOM-based calibration and the neural network approach described in this work.

\subsection{Kilo-Degree Survey}

We find more significant differences with respect to the published KiDS-1000 results. In the first tomographic bin, we find substantially more galaxies at high redshift. This leads to a shift in the mean redshift whereas we find fairly consistent results regarding the median, $0.229$ vs. $0.234$. Another noteworthy difference is the distribution in the third tomographic bin where the bulk of the $n(z)$ is shifted towards lower redshifts while the mean redshift is consistent due to an extended tail towards higher redshifts, similar to the first tomographic bin. Possible reasons for these discrepancies are discussed in the next section and the appendix.

\section{Discussion}
\label{sec:discussion}

\subsection{Direct Calibration}

In this paper, we have introduced a new technique for the direct calibration of photometric redshifts in weak lensing surveys. Whereas the fiducial studies all employ SOMs \citep{Hikage2019_PASJ_71_43, Hildebrandt2021_AA_647_124, Myles2021_MNRAS_505_4249}, our method utilizes neural network regression. Despite these apparent differences, conceptually, both methods share a number of similarities.

The deep-wide SOM approach developed in \cite{Buchs2019_MNRAS_489_820} uses two sets of photometric data and associated SOMs\footnote{HSC and KiDS do not explicitly use the deep-wide SOM approach and, instead, rely on a single SOM. Such a single SOM approach can be regarded as a special case of the deep-wide SOM method where wide and deep SOM and associated photometry are identical.}. First, there are deep-photometry fields with more color information and higher depth than the wide weak lensing survey. These deep photometric fields include the calibration sample with high-quality redshifts. One then groups objects into galaxies occupying the same cell in the deep SOM. Within each SOM cell, the galaxies in the calibration sample are assumed to be representative of all galaxies, including those without high-quality redshifts. Effectively, galaxies in the calibration sample are upweighted by $N_\mathrm{SOM} / N_\mathrm{cal}$, where $N_\mathrm{SOM}$ is the number of galaxies in the SOM cell and $N_\mathrm{cal}$ the number of galaxies in the SOM cell that also have high-quality redshifts. The deep SOM thus serves the purpose of alleviating selection biases in the calibration sample and the similarities to the importance weights discussed in section \ref{subsec:incompleteness} should be apparent.

The deep-wide SOM approach also utilizes a SOM of the shallower wide photometry of the entire survey. In particular, the wide-photomery color distribution of galaxies in the deep field is compared to the wide-photomery color distribution of all galaxies. Effectively, galaxies in the deep fields are assigned additional weights such that their wide-photometry color distribution matches that of the entire survey. In principle, this step is not strictly necessary for unbiased results if the wide-photometry of deep field galaxies is representative. However, this may reduce cosmic variance, especially if colors and thereby wide-photometry photometric redshifts correlate strongly with high-quality redshifts within a tomographic bin. One way to implement this technique is to correct the mean redshift obtained from high-quality redshifts $\bar z$ for the difference between the mean photometric redshift of the deep fields $\bar z_\mathrm{p}$ and the entire survey using the control variates technique. This will reduce the variance in the mean redshift by a factor $\eta$, i.e., $\tilde \sigma_{\bar z}^2 = \eta \sigma_{\bar z}^2$, with
\begin{equation}
    \begin{split}
    \eta &= 1 - \frac{\mathrm{Cov}^2(\bar z, \bar z_\mathrm{p})}{\sigma_{\bar z}^2 \sigma_{\bar z_\mathrm{p}}^2}\\
    &\approx 1 - \frac{\left[ \frac{\mathrm{Cov}(z, z_\mathrm{p})}{N_\mathrm{eff}} + \sigma_{\bar z, \mathrm{CV}}^2 \right]^2}{\left( \sigma_{\bar z, \mathrm{SN}}^2 + \sigma_{\bar z, \mathrm{CV}}^2 \right) \left( \sigma_{\bar z_\mathrm{p}, \mathrm{SN}}^2 + \sigma_{\bar z, \mathrm{CV}}^2 \right)} \, .
    \end{split}
\end{equation}
In the last transformation, we have optimistically assumed that changes in the mean redshift due to cosmic variance are, on average, perfectly reproduced by the photometric reshifts. Even in this optimistic scenario, we find $\eta \sim 0.5$ for all bins of HSC and KiDS, indicating that the weighting by the wide SOM would not dramatically increase the precision of our estimates. For this reason, our method does not currently implement an equivalent of weighting by the wide SOM distribution.

One complication of the SOM method is the inherent pixelization, i.e., galaxies are uniquely associated with individual cells or pixels in the SOM. The number of galaxies associated with each SOM cell can be highly variable, especially for the smaller sample of galaxies with high-quality redshifts. However, the SOM method requires every SOM cell to be occupied, thus necessitating a trade-off between the resolution of the SOM, i.e., the number of pixels it has, and the requirement to have at least one galaxy in each SOM cell.\footnote{In the KiDS-1000 analysis, some SOM cells were allowed to be unoccupied. However, cells were later grouped into a number of clusters that were then treated like a single cell.} Our method based on neural networks avoids this issue altogether by not requiring any pixelization. Additionally, we established a simple one-parameter ($p_\mathrm{f, min}$) method to adjust how much the calibration relies on high-quality photometric redshifts for galaxies insufficiently covered by spectroscopic samples.

\subsection{KiDS-1000 Results}

Whereas our results for DES Y3 and HSC Y1 are in good agreement with the fiducial results presented in \cite{Myles2021_MNRAS_505_4249} and \cite{Hikage2019_PASJ_71_43}, we find stronger differences with the KiDS-1000 results presented in \cite{Hildebrandt2021_AA_647_124}, as indicated in Fig.~\ref{fig:n_z} and Table~\ref{tab:n_z}.

One major difference in our analysis is the use of high-quality photometric redshifts from the COSMOS2020 catalog to supplement spectroscopic data. Using COSMOS2020 redshifts is necessary since some galaxies are severely under-represented in the DESI COSMOS-XMM catalog. The effect of including COSMOS2015 multi-band redshifts in the KiDS-1000 analysis was studied in \cite{vandenBusch2022_AA_664_170} using the same methodology as in \cite{Hildebrandt2021_AA_647_124}. It was found that a calibration that includes COSMOS2015 redshifts leads to higher mean redshifts even for identical SOM cells. Since we also find higher mean redshifts on average, this indicates that the use of COSMOS2020 redshifts may be one of the reasons for the different results we find here. In turn, part of this may be driven by scatter and catastrophic outliers in COSMOS2020 redshifts \citep{Weaver2022_ApJS_258_11}.

As mentioned earlier, in principle, assigning extra weights to galaxies in the calibration field to reproduce the color and magnitude distribution in the entire survey is not necessary for unbiased results. This requires that the color distribution and depth of the calibration field is representative of the full survey. However, for KiDS-1000, we find that the KiDS observations in the COSMOS field are somewhat deeper than the wider survey. Whereas \cite{Hildebrandt2021_AA_647_124} implicitly make the KiDS COSMOS catalog reproduce the color and magnitude distribution in KiDS-1000 via a SOM, we did not. This may also explain some of the differences in our results. However, it is not clear that applying extra weights, as was done in \cite{Hildebrandt2021_AA_647_124}, resolves all issues associated with non-representative photometry. In the appendix, we present a comparison of the color and magnitude distribution of galaxies in the KiDS COSMOS catalog compared to the wider survey.

Finally, another reason for the different results may be related to the SOM cell quality cuts that were applied in the KiDS-1000 analysis \citep{Hildebrandt2021_AA_647_124}. Before performing the photometric redshift calibration, \cite{Hildebrandt2021_AA_647_124} defined the so-called gold selection by removing galaxies for which the mean spectroscopic redshifts of the calibration sources differ systematically from the mean photometric redshifts in that SOM cell.
\begin{equation}
    \begin{split}
        \left| \langle z \rangle_\mathrm{d} - \langle z_\mathrm{p} \rangle_\mathrm{w} \right| <
        \max \left[ 5 \, \mathrm{nMAD} \left( \langle z \rangle_\mathrm{d} - \langle z_\mathrm{p} \rangle_\mathrm{d} \right), 0.4 \right]
    \end{split}
\end{equation}
In the above equation, the subscripts d and w denote averages of the deep calibration field and the wide survey, respectively. Also, whereas the quality cut is applied to each SOM cell individually, the normalized median absolute deviation on the right-hand side is taken over all SOM cells. The goal of this quality cut is to remove parts of the color and magnitude space for which photometric redshifts are  less reliable. Using the MICE mock catalogs, \cite{Wright2020_AA_637_100} show that this is indeed the case and reduces the bias of the SOM direct calibration method.

These advantages notwithstanding, the quality cut may lead to slight statistical biases. Whether any SOM cell fulfills this quality cut is a random variable and can be very stochastic given that one typically has $\mathcal{O}(10)$ galaxies with spectroscopic redshifts in each SOM cell. A potential issue is that \cite{Hildebrandt2021_AA_647_124} use the same galaxies to define the quality cut and to perform the photometric redshift calibration. In particular, by construction, all SOM cells in the gold selection pass the quality cut, i.e., do not show significant differences between the spectroscopic and photometric redshifts. For example, the analysis of \cite{Hildebrandt2021_AA_647_124} finds that for the first tomographic bin, KiDS-1, no single SOM cell has a mean redshift above $\sim 0.6$ which implies that the reconstructed $n(z)$ drops off sharply for $z \gtrsim 0.6$. However, this is, to some extent, a statistical artifact from the fact that the same sample was used to determine the quality cuts and to perform the calibration. If one uses galaxies not used to define the quality cut, as was done in this work, it is possible that individual SOM cells may have mean redshifts above $\sim 0.6$. This will lead to a higher $n(z)$ at high redshifts, as we see in the comparison of the reconstructed $n(z)$ for KiDS-1. In summary, the implementation of the quality cuts in \cite{Hildebrandt2021_AA_647_124} may introduce subtle statistical biases which may also be partially responsible for the different results.

\begin{figure}
    \centering
    \includegraphics{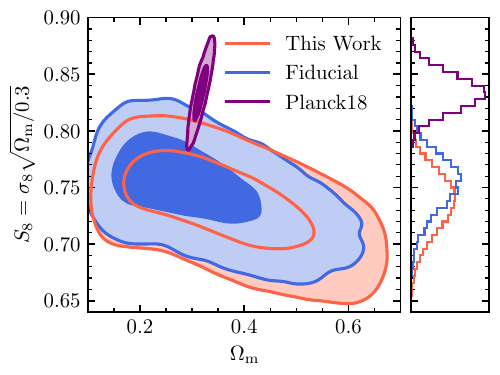}
    \caption{Posterior constraints on $S_8$ and $\Omega_\mathrm{m}$ from the KiDS-1000 cosmic shear analysis of COSEBIs. In blue, we show the fiducial results from \protect\cite{Asgari2021_AA_645_104}, whereas we show the updated results using the redshift distribution derived in this work in red. We also show the \cite{PlanckCollaboration2020_AA_641_6} TT,TE,EE+lowE results in purple. The two-dimensional histogram shows the $68\%$ and $99\%$ containment ranges and has been smoothed for visual clarity.}
    \label{fig:s_8}
\end{figure}

Despite these differences in the derived redshift distributions, we do not find significant changes in the cosmological constraints from cosmic shear. In Fig.~\ref{fig:s_8}, we show the posterior constraints on $\Omega_\mathrm{m}$, the matter density, and $S_8 = \sigma_8 \sqrt{\Omega_\mathrm{m} / 0.3}$, where $\sigma_8$ is the amplitude of matter fluctuations. The results come from a re-analysis of COSEBIs (Complete Orthogonal Sets of E/B-Integrals) in the KiDS-1000 cosmic shear data set using the CosmosSIS pipeline \citep{Zuntz2015_AC_12_45}, as described in \cite{Asgari2021_AA_645_104}. The only noteworthy difference is that we use the {\sc nautilus} Bayesian sampler \citep{Lange2023_MNRAS_525_3181} instead of {\sc MultiNest} \citep{Feroz2009_MNRAS_398_1601} to compute the posterior, leading to more accurate but slightly weaker posterior constraints \citep{Lemos2023_MNRAS_521_1184}. In Fig.~\ref{fig:s_8}, we compare the fiducial results using the redshift calibration of \cite{Hildebrandt2021_AA_647_124} with those derived from our updated photometric redshift calibration. Note that the original cosmic shear analysis of \cite{Asgari2021_AA_645_104} assumes that the redshift calibration may be biased in the mean redshift with Gaussian priors on the offset. Whereas the fiducial result uses priors with non-zero offsets derived from MICE mock catalogs, as described in \cite{Hildebrandt2021_AA_647_124}, the calibration offset priors are zero-centered for our updated analysis. Overall, we do not find strong differences in the derived constraints. While the derived constraint on $S_8$ is lower by $\sim 0.01$, the change happens mostly along the $S_8$-$\Omega_\mathrm{m}$-degeneracy direction and, thus, has little impact on the tension between weak gravitational lensing and the \cite{PlanckCollaboration2020_AA_641_6} CMB results, the so-called $S_8$-tension \citep{Abdalla2022_JHEAp_34_49}.

\subsection{DESI COSMOS-XMM}

In this work, we have presented the first unified photometric redshift calibration of weak lensing data with DESI redshifts. We find that the single DESI data set is already competitive with respect to the entirety of previous spectroscopic redshift data sets combined. For example, the DES Y3 analysis by \cite{Myles2021_MNRAS_505_4249} utilized spectroscopic redshifts from zCOSMOS \citep{Lilly2009_ApJS_184_218}, C3R2 \citep{Masters2019_ApJ_877_81}, VVDS \citep{LeFevre2013_AA_559_14}, and VIPERS \citep{Scodeggio2018_AA_609_84} and was able to calibrate up to $47\%$ of the weak lensing source redshift distribution with spectroscopic redshifts. In contrast, in this analysis for $p_\mathrm{f, min} = 0.1$, we are able to calibrate $69\%$ of DES with DESI redshifts alone. Subsequent iterations of the DESI COSMOS-XMM catalogs will likely increase this fraction further, making DESI an ideal data set for redshift calibration. In particular, our neural network fiber assignment probabilities allow us to easily identify bright, easy-to-observe galaxies that are currently undersampled in DESI COSMOS-XMM. At the same time, DESI's coverage is currently poor for fainter source galaxies, as indicated by the much lower spectroscopic fractions for the HSC samples. Additionally, near-IR follow-up observations may be needed for high-redshift objects for which redshifts are hard to obtain within DESI's wavelength coverage, $360$ nm to $980$ nm. Thus, calibrating a significant fraction of weak lensing source galaxies from the Vera C. Rubin Observatory down to faint magnitudes would require additional deep observations. In the future, we plan to supplement DESI redshifts with other spectroscopic samples such as zCOSMOS. Finally, we want to point out that the footprint of the DESI COSMOS-XMM catalog extends far beyond the original, roughly $1 \, \mathrm{deg}^2$ COSMOS footprint and the corresponding DES, HSC, and KiDS COSMOS footprints, as shown in Fig.~\ref{fig:map}. Thus, extending photometric data to better match the DESI COSMOS-XMM spectroscopy footprint, as is expected with the Rubin deep drilling fields \citep{Scolnic2018_arXiv_1812_0516}, will lead to a substantial reduction in cosmic variance and shape noise.

\section{Conclusion}
\label{sec:conclusion}

In this work, we have presented the first unified photometric redshift calibration of the leading stage-III weak lensing surveys: DES, HSC, and KiDS. Furthermore, we have performed the first weak lensing source redshift calibration with the DESI COSMOS-XMM catalogs. Finally, we also introduced a new method for direct redshift calibration using neural networks. This method represents an alternative to the widely-used SOM direct calibration method and, among others, avoids the pixelization issue inherent in the latter. Our main findings when applying this method to data are as follows:
\begin{itemize}
    \item Our neural network approach is able to model the complex selection function inherent in the DESI COSMOS-XMM catalog. Additionally, we showed that our redshift calibration method can recover intrinsic redshift distributions from this incomplete, biased sample.
    \item We find excellent agreement with previously published redshift calibration results for DES Y3 \citep{Myles2021_MNRAS_505_4249} and HSC Y1 \citep{Hikage2019_PASJ_71_43}. On the other hand, more significant differences exists with respect to the results of the fiducial KiDS-1000 analysis \citep{Wright2020_AA_637_100, Hildebrandt2021_AA_647_124}. We speculate that this difference may be related the use of COSMOS2020 multi-band redshifts, the peculiar photometry in the KiDS COSMOS field, and subtle statistical biases in the use of quality cuts.
    \item Our results have a high statistical precision with $\sigma_{\bar z} < 0.01$ for most samples. Additionally, the DESI sample covers a large fraction of the color and magnitude space, allowing use to calibrate the majority of the DES and KiDS weak lensing sample with DESI spectroscopic redshifts alone.
\end{itemize}
Our analysis method and data are also readily applicable to other weak lensing data sets such as the weak lensing catalogs from the Sloan Digital Sky Survey \citep[SDSS; ][]{Mandelbaum2005_MNRAS_361_1287, Mandelbaum2013_MNRAS_432_1544} or the Dark Energy Camera Legacy Survey \citep[DECaLS; ][]{Yao2023_MNRAS_524_6071}. Most importantly, we find that DESI is an excellent instrument for the photometric redshift calibration of future data sets such as the weak lensing catalog from the Vera C. Rubin Observatory. To fully achieve this potential, we also showed that larger-area deep fields matching the DESI focal plane footprint are needed.

Recent cosmological tensions such as the $S_8$-tension \citep{Abdalla2022_JHEAp_34_49} between cosmic microwave background observations and weak lensing measurements sensitively depend on redshift calibration. For the current stage-III surveys, our analysis suggests that redshift calibration is robust enough as to not strongly affect cosmology results. In the future, DESI may play a crucial role in achieving the stringent calibration requirements of stage-IV weak lensing surveys.

\section*{Acknowledgments}

We thank Justin Myles and Biprateep Dey for comments on the manuscript. We are also grateful to the anonymous referee whose detailed and thorough report improved the clarity of the paper.

This work made use of the following software: {\sc Astropy} \citep{AstropyCollaboration2013_AA_558_33}, {\sc matplotlib} \citep{Hunter2007_CSE_9_90}, {\sc NumPy} \citep{VanDerWalt2011_CSE_13_22}, {\sc SciPy} \citep{Virtanen2020_NatMe_17_261}, {\sc scikit-learn} \citep{Pedregosa2011_JMLR_12_2825}, {\sc CAMB} \citep{Lewis2000_ApJ_538_473, Howlett2012_JCAP_04_027}, and {\sc Spyder}.

AHW and HHi are supported by funding from the German Science Foundation DFG, via the Collaborative Research Center SFB1491 “Cosmic Interacting Matters - From Source to Signal”. AHW is also supported by the Deutsches Zentrum für Luft- und Raumfahrt (DLR), made possible by the Bundesministerium für Wirtschaft und Klimaschutz. HHi is also supported by a Heisenberg grant of the Deutsche Forschungsgemeinshaft (Hi 1495/5-1), and a European Research Council Consolidator Grant (No. 770935). Work by AF has been supported by a CAS Summer Undergraduate Fellowship from American University’s College of Arts \& Sciences (CAS) and the NASA DC Space Grant Consortium (DCSGC).

This material is based upon work supported by the U.S. Department of Energy (DOE), Office of Science, Office of High-Energy Physics, under Contract No. DE–AC02–05CH11231, and by the National Energy Research Scientific Computing Center, a DOE Office of Science User Facility under the same contract. Additional support for DESI was provided by the U.S. National Science Foundation (NSF), Division of Astronomical Sciences under Contract No. AST-0950945 to the NSF’s National Optical-Infrared Astronomy Research Laboratory; the Science and Technology Facilities Council of the United Kingdom; the Gordon and Betty Moore Foundation; the Heising-Simons Foundation; the French Alternative Energies and Atomic Energy Commission (CEA); the National Council of Humanities, Science and Technology of Mexico (CONAHCYT); the Ministry of Science, Innovation and Universities of Spain (MICIU/AEI/10.13039/501100011033), and by the DESI Member Institutions: \url{https://www.desi.lbl.gov/collaborating-institutions}. Any opinions, findings, and conclusions or recommendations expressed in this material are those of the author(s) and do not necessarily reflect the views of the U. S. National Science Foundation, the U. S. Department of Energy, or any of the listed funding agencies.

The authors are honored to be permitted to conduct scientific research on I'oligam Du'ag (Kitt Peak), a mountain with particular significance to the Tohono O’odham Nation.

\textbf{Dark Energy Survey:} This project used public archival data from the Dark Energy Survey (DES). Funding for the DES Projects has been provided by the U.S. Department of Energy, the U.S. National Science Foundation, the Ministry of Science and Education of Spain, the Science and Technology FacilitiesCouncil of the United Kingdom, the Higher Education Funding Council for England, the National Center for Supercomputing Applications at the University of Illinois at Urbana-Champaign, the Kavli Institute of Cosmological Physics at the University of Chicago, the Center for Cosmology and Astro-Particle Physics at the Ohio State University, the Mitchell Institute for Fundamental Physics and Astronomy at Texas A\&M University, Financiadora de Estudos e Projetos, Funda{\c c}{\~a}o Carlos Chagas Filho de Amparo {\`a} Pesquisa do Estado do Rio de Janeiro, Conselho Nacional de Desenvolvimento Cient{\'i}fico e Tecnol{\'o}gico and the Minist{\'e}rio da Ci{\^e}ncia, Tecnologia e Inova{\c c}{\~a}o, the Deutsche Forschungsgemeinschaft, and the Collaborating Institutions in the Dark Energy Survey.

The Collaborating Institutions are Argonne National Laboratory, the University of California at Santa Cruz, the University of Cambridge, Centro de Investigaciones Energ{\'e}ticas, Medioambientales y Tecnol{\'o}gicas-Madrid, the University of Chicago, University College London, the DES-Brazil Consortium, the University of Edinburgh, the Eidgen{\"o}ssische Technische Hochschule (ETH) Z{\"u}rich,  Fermi National Accelerator Laboratory, the University of Illinois at Urbana-Champaign, the Institut de Ci{\`e}ncies de l'Espai (IEEC/CSIC), the Institut de F{\'i}sica d'Altes Energies, Lawrence Berkeley National Laboratory, the Ludwig-Maximilians Universit{\"a}t M{\"u}nchen and the associated Excellence Cluster Universe, the University of Michigan, the National Optical Astronomy Observatory, the University of Nottingham, The Ohio State University, the OzDES Membership Consortium, the University of Pennsylvania, the University of Portsmouth, SLAC National Accelerator Laboratory, Stanford University, the University of Sussex, and Texas A\&M University.

Based in part on observations at Cerro Tololo Inter-American Observatory, National Optical Astronomy Observatory, which is operated by the Association of Universities for Research in Astronomy (AURA) under a cooperative agreement with the National Science Foundation.

\textbf{Hyper Suprime-Cam:} The Hyper Suprime-Cam (HSC) collaboration includes the astronomical communities of Japan and Taiwan, and Princeton University. The HSC instrumentation and software were developed by the National Astronomical Observatory of Japan (NAOJ), the Kavli Institute for the Physics and Mathematics of the Universe (Kavli IPMU), the University of Tokyo, the High Energy Accelerator Research Organization (KEK), the Academia Sinica Institute for Astronomy and Astrophysics in Taiwan (ASIAA), and Princeton University. Funding was contributed by the FIRST program from the Japanese Cabinet Office, the Ministry of Education, Culture, Sports, Science and Technology (MEXT), the Japan Society for the Promotion of Science (JSPS), Japan Science and Technology Agency (JST), the Toray Science Foundation, NAOJ, Kavli IPMU, KEK, ASIAA, and Princeton University. 

This paper makes use of software developed for Vera C. Rubin Observatory. We thank the Rubin Observatory for making their code available as free software at \url{http://pipelines.lsst.io/}.

This paper is based on data collected at the Subaru Telescope and retrieved from the HSC data archive system, which is operated by the Subaru Telescope and Astronomy Data Center (ADC) at NAOJ. Data analysis was in part carried out with the cooperation of Center for Computational Astrophysics (CfCA), NAOJ. We are honored and grateful for the opportunity of observing the Universe from Maunakea, which has the cultural, historical and natural significance in Hawaii. 

\textbf{Kilo Degree-Survey:} Based on observations made with ESO Telescopes at the La Silla Paranal Observatory under programme IDs 177.A-3016, 177.A-3017, 177.A-3018 and 179.A-2004, and on data products produced by the KiDS consortium. The KiDS production team acknowledges support from: Deutsche Forschungsgemeinschaft, ERC, NOVA and NWO-M grants; Target; the University of Padova, and the University Federico II (Naples).

We use the gold sample of weak lensing and photometric redshift measurements from the fourth data release of the Kilo-Degree Survey \citep{Kuijken2019_AA_625_2, Wright2020_AA_637_100, Hildebrandt2021_AA_647_124, Giblin2021_AA_645_105}, hereafter referred to as KiDS-1000. Cosmological parameter constraints from KiDS-1000 have been presented in \cite[][cosmic shear]{Asgari2021_AA_645_104}, \cite[][$3\times2$pt]{Heymans2021_AA_646_140} and \cite[][beyond $\Lambda$CDM]{Troster2021_AA_649_88}, with the methodology presented in \cite{Joachimi2021_AA_646_129}.

\section*{Data Availability}

The majority of data used in this work is publicly available. The compilation of DESI redshifts in the COSMOS and XMM fields has been released at \url{https://data.desi.lbl.gov/public/papers/c3/cosmos-xmmlss/}. The COSMOS2020 photometric catalog is  available at \url{https://cosmos2020.calet.org/}. For DES Y3, the Balrog photometric catalogs can be accessed at \url{https://des.ncsa.illinois.edu/releases/y3a2/Y3balrog} with the {\sc METACAL} tables used for weak lensing cuts and weights available upon request from the DES collaboration. The DES Y3 fiducial redshift calibration can be retrieved from \url{https://des.ncsa.illinois.edu/releases/y3a2/Y3key-products}. The HSC Y1 photometric catalog in the COSMOS field can be found at \url{https://hsc-release.mtk.nao.ac.jp/doc/index.php/s17a-wide-cosmos/}. Finally, the KiDS-1000 photometric catalogs, weak lensing catalogs, and redshift calibration are available at \url{https://kids.strw.leidenuniv.nl/DR4/access.php}, \url{https://kids.strw.leidenuniv.nl/DR4/KiDS-1000_shearcatalogue.php}, and \url{https://kids.strw.leidenuniv.nl/DR4/data_files/KiDS1000_SOM_N_of_Z.tar.gz}, respectively.

All data points shown in the published graph and codes used are available at \url{https://zenodo.org/doi/10.5281/zenodo.15164171}. All data products produced in this work are available upon reasonable request to the lead author.

\bibliography{bibliography}

\appendix

Our analysis method inherently assumes that the properties of galaxies, in the deep fields, including their measured photometry, are representative of the wider weak lensing survey. However, this may not be the case for KiDS-1000. One reason for this is that the KiDS pointings have a range of depths in the $ugri$ bands and the COSMOS pointing may not perfectly representative of the mean. For example, while the $r$-band depth of COSMOS pointing is within the range of KiDS-1000 pointings, it is deeper than the average. On the other hand, the $ZYJHK_s$ bands were actually observed at higher depth but artificially downgraded by adding gaussian noise. Ultimately, for KiDS-1000, we find statistically significant differences in the color and magnitude distribution of KiDS observations in COSMOS deep field compared to KiDS-1000. 

\begin{figure*}
    \centering
    \includegraphics{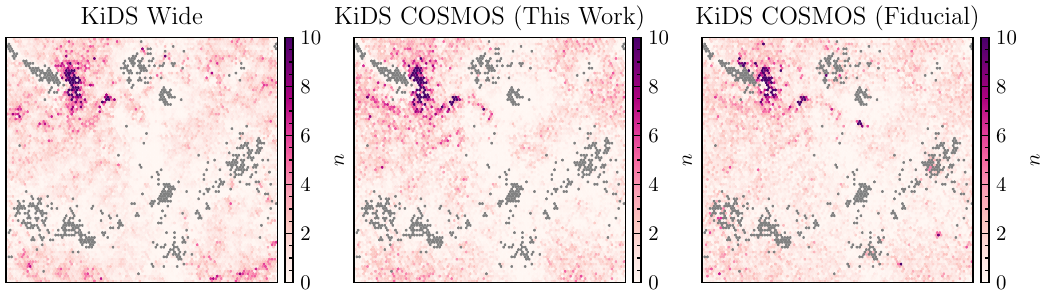}
    \caption{The effective number of galaxies in each SOM cell for the full KiDS SOM-gold sample (left), the transfer catalog used in this work (middle), and the transfer catalog used in \protect\cite{Hildebrandt2021_AA_647_124}. The values are normalized such that the average number over all SOM cells is 1 in all cases. Neither transfer catalog faithfully reproduces all features of the KiDS SOM-gold catalog.}
    \label{fig:kids_transfer_som}
\end{figure*}

Fig.~\ref{fig:kids_transfer_som} shows the color distribution of galaxies in the KiDS SOM, weighted by the lensing weight $w_\mathrm{l}$. We compare the results from KiDS-1000 against our COSMOS catalog using the weights derived from neural networks and the fiducial KiDS COSMOS catalog using weights derived from nearest-neighbor matching \citep{Hildebrandt2021_AA_647_124}. While the KiDS COSMOS catalog reproduce many salient features in the SOM map, clear differences also emerge, regardless of the assumed weight in KiDS COSMOS. For example, one can see that the distribution of galaxies in the KiDS COSMOS catalog reaches to fainter $r$-band magnitudes than the KiDS-1000 gold catalog.

\begin{figure}
    \centering
    \includegraphics{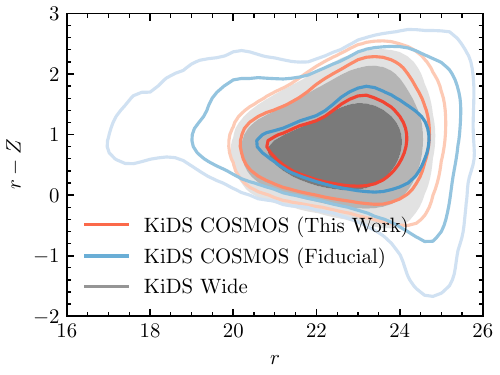}
    \caption{Weighted distribution of $r$ magnitudes and $r-Z$ colors of galaxies in the KiDS SOM-gold sample (grey) compared to the KiDS COSMOS catalog used in this work (red) and the fiducial KiDS COSMOS catalog (blue). We show $99\%$, $95\%$, and $68\%$ ranges. The densities have been smoothed with a Gaussian kernel with a width of $0.2$ mag for visual clarity.}
    \label{fig:kids_transfer_magnitudes}
\end{figure}

This is further investigated in Fig.~\ref{fig:kids_transfer_magnitudes}, where we show the corresponding $r$ vs. $r-Z$ color and magnitude distributions. The KiDS COSMOS catalog using nearest-neighbor weights clearly shows substantial differences to KiDS-1000. The main reason is that these weights do not account for the probability $w_\mathrm{t}(s)$ that a galaxy makes the KiDS-1000 weak lensing cuts. However, even for the neural network weights that account for $w_\mathrm{t}(s)$, the KiDS COSMOS catalog extends to deeper magnitudes and shows redder colors than KiDS-1000. We observe similar color and magnitude trends in the raw catalogs, i.e., before applying lensing weights, indicating that this is not due to our approximate method for creating KiDS-1000-like lensing weights. All of this indicates that the KiDS COSMOS catalog is not perfectly representative of the KiDS-1000 data set and may lead to biases in the estimated redshift distribution.

\end{document}